\documentclass[10pt,journal]{IEEEtran}
\ifCLASSOPTIONcompsoc
\usepackage[nocompress]{cite}
\else
\usepackage{cite}
\fi

\ifCLASSINFOpdf
    \usepackage[pdftex]{graphicx}
    \graphicspath{{../figures/}}
    \DeclareGraphicsExtensions{.pdf,.jpeg,.png}
\else
\fi
\usepackage{amsmath}
\usepackage{algorithmic}

\usepackage{array}
\usepackage{url}

\usepackage{amsfonts}
\usepackage{amssymb}
\usepackage{amsthm}
\usepackage{enumerate}
\usepackage{float}
\usepackage{color}
\usepackage{stfloats,fancyhdr}
\usepackage{multirow}
\usepackage{changepage}
\usepackage[normalem]{ulem}

\usepackage{comment}

\usepackage{url}
\usepackage{hyperref}
\usepackage{mathtools}
\usepackage{amsmath}

\usepackage{flushend}

\newcommand{\FigPath}{./figures/}

\begin{document}
%
\title{Minimizing Age of Collection for Multiple Access in Wireless Industrial Internet of Things}
%
%
%

\author{Jiaxin~Liang,~\IEEEmembership{Member,~IEEE,}~Tse-Tin~Chan,~\IEEEmembership{Member,~IEEE,}~and~Haoyuan~Pan,~\IEEEmembership{Member,~IEEE}
\IEEEcompsocitemizethanks{
\IEEEcompsocthanksitem Jiaxin Liang is with the Computer Network and Protocol Research Laboratory, Huawei Technologies, Shenzhen, 518060, China (e-mail: {jiaxin.liang@huawei.com}). The work was done when he was with the Department of Information Engineering, The Chinese University of Hong Kong, Hong Kong SAR, China.
\IEEEcompsocthanksitem Tse-Tin Chan is with the Department of Mathematics and Information Technology, The Education University of Hong Kong, Hong Kong SAR, China (e-mail: {tsetinchan@eduhk.hk}).
\IEEEcompsocthanksitem Haoyuan Pan is with the College of Computer Science and Software Engineering, Shenzhen University, Shenzhen, 518060, China (e-mail: {hypan@szu.edu.cn}).

}
}
\maketitle

\begin{abstract}
This paper investigates the information freshness of Industrial Internet of Things (IIoT) systems, where each IoT device makes a partial observation of a common target and transmits the information update to a central receiver to recover the complete observation. We consider the age of collection (AoC) performance as a measure of information freshness. Unlike the conventional age of information (AoI) metric, the instantaneous AoC decreases only when all cooperative packets for a common observation are successfully received. Hence, effectively allocating wireless time-frequency resources among IoT devices to achieve a low average AoC at the central receiver is paramount. Three multiple access schemes are considered in this paper: time-division multiple access (TDMA) without retransmission, TDMA with retransmission, and frequency-division multiple access (FDMA). First, our theoretical analysis indicates that TDMA with retransmission outperforms the other two schemes in terms of average AoC. Subsequently, we implement information update systems based on the three schemes on software-defined radios. Experimental results demonstrate that considering the medium access control (MAC) overhead in practice, FDMA achieves a lower average AoC than TDMA with or without retransmission in the high signal-to-noise ratio (SNR) regime. In contrast, TDMA with retransmission provides a stable and relatively low average AoC over a wide SNR range, which is favorable for IIoT applications. Overall, we present a theoretical-plus-experimental investigation of AoC in IIoT information update systems.
\end{abstract}


%
\IEEEpeerreviewmaketitle

\section{Introduction}
%
%
%
%
\IEEEPARstart{T}{he} Industrial Internet of Things (IIoT), a key enabler for future inter-device communication in industrial development, has received much attention in the past few years. IIoT refers to networks of sensors, cameras, controllers, and other devices used to monitor and control industrial processes. Several metrics have been proposed for consideration when designing networks for IIoT, such as throughput, reliability, and latency. Recently, research has paid significant attention to a new perspective on the IIoT: information freshness. The freshness of information updates is vital for industrial automation applications involving time-sensitive measurements and real-time control~\cite{On the role of age of information in the Internet of Things}, such as autonomous vehicles, digital twins, tactile Internet, and smart cities, because outdated information can lead to incorrect decisions.

To quantify the information freshness at the destination, a new performance metric, Age of Information (AoI), has gained considerable interest in recent years~\cite{Age of information: An introduction and survey}. AoI characterizes the time interval from the generation of the last successfully received packet to the current time~\cite{Age of information: A new concept}. Distinguishing from conventional metrics such as throughput, reliability, and latency, AoI jointly characterizes 1) the interval at which information updates are generated and 2) the end-to-end transmission delay in communication. Therefore, from the destination's perspective, AoI is more comprehensive and useful for assessing the freshness of the latest received packet~\cite{Age of Information: Foundations and Applications}. Numerous works, such as~\cite{Age of information with hybrid-ARQ: A unified explicit result, Power minimization for age of information constrained dynamic control in wireless sensor networks, Optimizing information freshness for cooperative IoT systems with stochastic arrivals, Centralized and distributed age of information minimization with nonlinear aging functions in the Internet of things, A reinforcement learning approach for optimizing the age-of-computing-enabled IoT}, have shown that replacing those conventional performance metrics with AoI leads to fundamental changes in system design.

In addition to AoI, different related performance metrics have been proposed to address application-specific requirements, such as value of information (VoI)~\cite{Age-of-information vs. value-of-information scheduling for cellular networked control systems, Optimal information updating based on value of information, Value of information in wireless sensor network applications and the IoT: A review}, quality of information (QoI)~\cite{Not just age but age and quality of information, Age of information for updates with distortion: Constant and age-dependent distortion constraints}, age of incorrect information (AoII)~\cite{The age of incorrect information: A new performance metric for status updates, The age of incorrect information: An enabler of semantics-empowered communication, Age of incorrect information for remote estimation of a binary Markov source}, etc. However, most existing AoI or AoI-related works assume that different devices are associated with independent observations. When a packet from a device arrives at its destination, the knowledge of the corresponding observation is updated. In practice, many IIoT applications involve monitoring common or related information by multiple devices, where it may not be appropriate to consider the AoI of each source independently. Therefore, recent research has started to address this issue for the development of IIoT communications. In the following, we discuss three groups of related works that jointly consider the AoI of multiple devices.

The first group considered multiple devices in which each observes the same source, for example, by placing redundant sensors to monitor the same physical phenomenon reliably. In other words, the AoI at the destination is reduced when a fresh information update from any of the devices observing the same source is successfully delivered. Reference~\cite{Minimizing the age of information from sensors with common observations} considered the case where each sensor can monitor multiple sources, and the update of each source is observed by a random subset of the sensors. The authors proposed scheduling policies to minimize the average AoI. In~\cite{Timely monitoring of dynamic sources with observations from multiple wireless sensors}, the authors further investigated scheduling schemes, which minimize the average AoI, when observations are generated at will and the probability that a sensor observes a source depends on the state of that source.

The second group examined multiple devices observing different sources, where status updates from one source may contribute to providing (partial) information from other relevant sources. For example, IoT devices monitor different physical processes that have overlapping regions. Reference~\cite{Age-oriented scheduling of correlated sources in multi-server system} considered the case where a packet from one device has the chance to bring complete update information about other devices. The authors then proposed a queue-based scheduling policy to reduce the total AoI of all sources. In~\cite{Age-of-information oriented scheduling for multichannel IoT systems with correlated sources}, the authors considered that when a device succeeds in updating, the AoI of other devices that overlap with the monitoring area of that device is partially reduced. They then formulated the problem as a correlated restless multi-armed bandit (CRMAB) problem to minimize the average AoI of the system. Considering energy-constrained devices, \cite{Using correlated information to extend device lifetime} utilized fresher updates from correlated sources to extend the time between successive updates of the sources, thus increasing the devices' lifetime.

The last group considered that a group of devices cooperates to monitor a target, and each device is responsible for a part of the source. A successful information update occurs only when all  the cooperative packets are decoded. For example, multiple cameras monitor the same scene from different angles and transmit their captured images to a remote center, reconstructing a complete 3D image based on the received images~\cite{Minimizing age of correlated information for wireless camera networks}. 

To distinguish from AoI, different new terms such as the age of collection (AoC) or the age of correlated information (AoCI) are used to describe that the ``age'' is reduced when all relevant information for a common observation is successfully updated. Reference \cite{Minimizing age of correlated information for wireless camera networks, Joint assignment and scheduling for minimizing age of correlated information} studied the joint optimization of serving node assignment and transmission scheduling to minimize the peak AoCI. In contrast to \cite{Minimizing age of correlated information for wireless camera networks, Joint assignment and scheduling for minimizing age of correlated information}, the authors of \cite{On the age of information in Internet of Things systems with correlated devices} further considered that the status updates could arrive randomly or be generated at will. There were some studies \cite{Application-oriented scheduling for optimizing the age of correlated information: A deep-reinforcement-learning-based approach, Status update for correlated energy harvesting sensors: A deep reinforcement learning approach} using deep reinforcement learning (DRL) to minimize the average AoCI. Reference \cite{Application-oriented scheduling for optimizing the age of correlated information: A deep-reinforcement-learning-based approach} generalized the problems in \cite{Minimizing age of correlated information for wireless camera networks, Joint assignment and scheduling for minimizing age of correlated information} as establishing many-to-many associations between devices and applications through an IoT hub and developed a DRL-based scheduling scheme. Considering energy-constrained IoT devices, \cite{Status update for correlated energy harvesting sensors: A deep reinforcement learning approach} proposed a DRL-based status update algorithm for IoT networks with multiple correlated energy harvesting sensors. Recently, the AoC performance of a base station collecting information updates from spatially distributed nodes in the presence of wireless interference, when the locations of nodes and interferers are known or unknown, was analyzed in \cite{Age of broadcast and collection in spatially distributed wireless networks}.

In this paper, we focus on the scenario considered in the last group of works mentioned above. The IIoT applications we study aggregate related information from multiple sources to a central controller for analysis, processing, or decision-making. Since AoI characterizes the information freshness of packets from respective sources, it does not accurately describe the age changes in the scenario considered in this paper. To address this issue, this paper adopts the average AoC performance~\cite{Age of broadcast and collection in spatially distributed wireless networks}. The AoC decreases only when all collaborating IoT devices for monitoring a common target update successfully. The new rule on age updates in AoC makes existing schemes in AoI-based studies potentially inapplicable. In particular, we are concerned with the fundamental problem of how to divide the time-frequency resources of the wireless medium to achieve a low average AoC as a reference for IIoT development.

Orthogonal multiple access (OMA) schemes, such as time-division multiple access (TDMA) and frequency-division multiple access (FDMA), are common in many IoT systems due to the simplicity of implementation. A previous study~\cite{Information update: TDMA or FDMA?} revealed that TDMA systems have a much better AoI performance than FDMA systems. Due to the smaller allocated bandwidth, FDMA users take longer to transmit a sample. Hence, the update received at the end of transmission was generated a longer time ago, resulting in poor AoI performance~\cite{Information update: TDMA or FDMA?}. However, it is not obvious whether TDMA is better than FDMA in terms of AoC performance because the new update rule for age complicates the problem. Allocating time-frequency resources to minimize the average AoC of the system remains an open question.

\begin{figure}[t]
    \centering
    \includegraphics[width=3.5in]{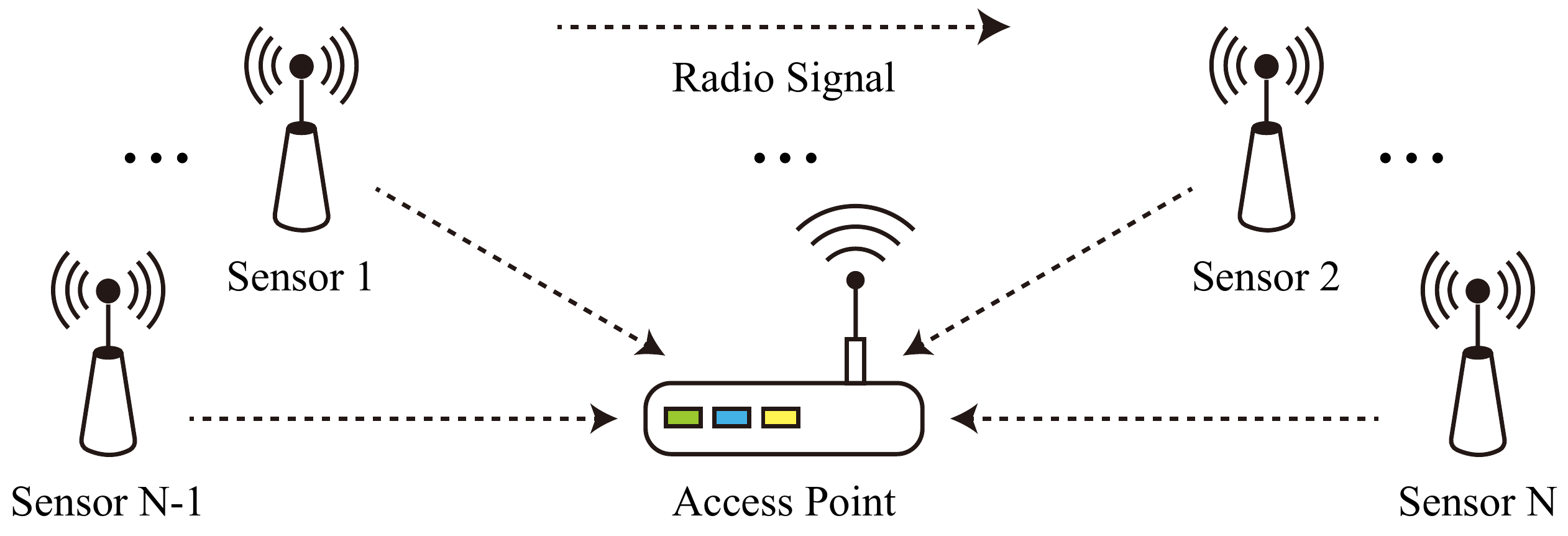}
    \caption{A wireless IIoT network with $N$ devices (e.g., sensors) observes common phenomena and reports them to an AP.}
    \label{fig:sys_arch}
\end{figure}

To fill this gap, this paper considers a general IIoT system where each IoT device partially observes a common target and transmits time-sensitive cooperative update packets to a common destination for recovering the complete observation, as shown in Fig. \ref{fig:sys_arch}. In particular, we investigate the transmission scheduling problem and study the average AoC of TDMA and FDMA systems. In TDMA systems, packets are transmitted one after another; thus, packet transmission failures can be reacted quickly to reduce the average AoC. Based on the different ways of using transmission feedback, two different TDMA schemes are considered: without retransmission and with retransmission. For TDMA without retransmission, all sent cooperative packets are discarded if any packet fails to be transmitted. In contrast, for TDMA with retransmission, the failed packet will be retransmitted without affecting other packets.

In FDMA systems, all packets from collaborating devices arrive at the central receiver simultaneously, which is naturally in the spirit of AoC: the information is updated after all cooperative packets are received. Unlike TDMA, FDMA does not need to wait for the transmission of some other packets, because all cooperative packets are transmitted simultaneously. However, if at least one packet from the collaborating devices fails to be decoded in FDMA, all the cooperative packets need to be discarded and wasted. Although FDMA with retransmission can be investigated, this requires reallocation of bandwidth based on the number of retransmitting devices, which complicates the design of IoT systems. In addition, if the IoT devices have a fixed transmit power, the effective signal-to-noise ratio (SNR) of the devices in retransmission decreases, resulting in a higher packet error rate (PER). Therefore, FDMA with retransmission is not considered in this paper.

This paper uses Markov chains to derive the theoretical AoC performance of FDMA and the two TDMA systems. We verify the theoretical AoC performance of the three multiple access schemes through MATLAB simulations. Our results show that TDMA with retransmission is theoretically superior to the other two schemes. To further explore the AoC performance in practical systems, we implement the three schemes on a Universal Software Radio Peripheral (USRP) software-defined Radio (SDR) platform. We design detailed medium access control (MAC) protocols for the three schemes, including designs of information update packet formats and link-layer feedback control mechanisms. Experimental results show that when MAC overhead is taken into account, FDMA can provide better average AoC performance than TDMA with retransmission at high SNRs. Overall, we provide a theoretical-plus-experimental investigation of AoC in IIoT information update systems.


\section{Age of Collection (AoC)}	\label{sec:preliminaries}

We study the AoC of a wireless IIoT network with $N$ IoT devices observing common phenomena and reporting the latest status to a central access point (AP), as shown in Fig. \ref{fig:sys_arch}. Specifically, we consider that each device cooperatively records partial information about the target at the same moment and sends an update packet to the AP in a multiple-access manner.

The complete observation is only available when all packets about the observation are received. Therefore, unlike most of the literature in which the AP measures the AoI of each device independently, the AoC considered in this paper focuses on the time elapsed until the AP receives updates from all devices. Specifically, the instantaneous AoC of the system at time $t$ is given by
\begin{equation}
    \Delta (t) = t - G(t)
\end{equation}
where $G(t)$ is the generation time of the most recently received status update packets from all the $N$ IoT devices. We remark that the instantaneous AoC $\Delta (t)$ drops only when all devices update successfully at the AP. This update rule differs from that in conventional AoI scenarios, where different devices are typically associated with independent observations---when an update packet from a particular device arrives, the instantaneous AoI of that device measured at the AP will drop independently. 

Similar to the definition of the average AoI, the average AoC, defined as the time average of the instantaneous AoC $\Delta (t)$, can be found by
\begin{equation} \label{eqa:general_avg_aoc}
    \bar \Delta  = \mathop {\lim }\limits_{T \to \infty } \frac{1}{T}\int_0^T {\Delta (t)\ dt}.
\end{equation}
A lower average AoC of the system indicates that the joint observations at the AP are generally fresher, better reflecting the global status measured by the devices. 

In multiple access systems, the wireless channel is shared among multiple devices. Different access schemes, such as TDMA and FDMA, study how to divide the channel among devices. In order to enable devices with correlated information (observed from the same phenomenon) to update the global status efficiently, we next examine the average AoC of systems using TDMA and FDMA.


\section{AoC of TDMA}
\label{AoC of TDMA}
\begin{figure}[t]
    \centering
    \includegraphics[width=3.5in]{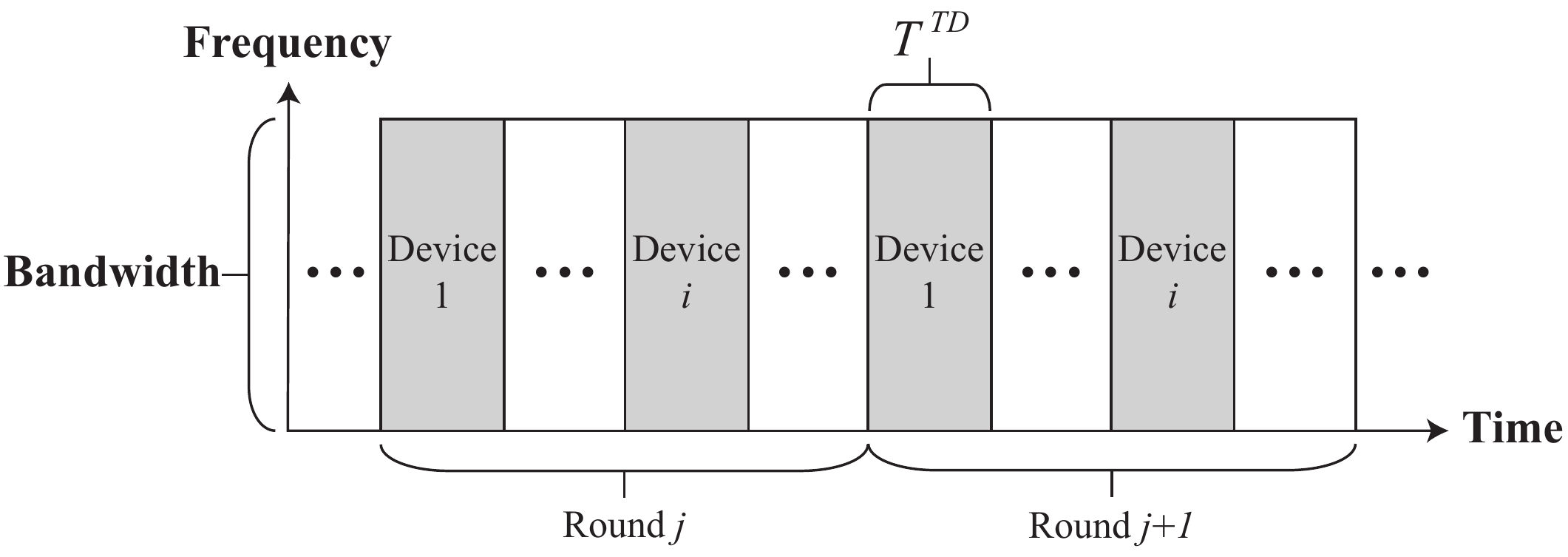}
    \caption{The TDMA scheme where devices take turns to access the channel.}
    \label{fig:tdma}
\end{figure}

We start our analysis with the AoC of the TDMA scheme. In a TDMA system, devices take turns sending their status packets to an AP in distinct time slots. We assume that the packets from all devices have the same block length $L$, and all time slots in the TDMA scheme have the same duration $T^{TD}$, as shown in Fig. \ref{fig:tdma}. The total bandwidth of the system is $B$. For ease of explanation, we use the number of time slots as the base unit of duration in the analysis of the TDMA schemes. In the final derivation of the results, we will convert this unit to seconds.

Recall that the AoC increases over time until the AP successfully collects the cooperative status packets from all devices, i.e., from the first to the last device in TDMA. A noteworthy question is, when a status packet transmission fails, should the failed device retransmit, or should all devices start a new transmission round (i.e., with or without retransmission)? A quantitative study is needed to study which one performs better than the other in terms of average AoC performance.

In the following, we first analyze the instantaneous AoC and the average AoC of a TDMA scheme without a retransmission strategy. We refer to the \textbf{TDMA} scheme with \textbf{N}o \textbf{R}etransmission strategy as \textit{\textbf{TDMA-NR}}. The subsequent subsection will analyze a TDMA scheme with a retransmission strategy.

\subsection{TDMA-NR}
In a TDMA-NR system, when the AP fails to decode the packet from device $i$ in round $j$, the AP immediately uses an acknowledgment (ACK) to notify all the devices of the failure, ask them to generate new samples, and coordinate them for the $(j+1)$-th round of transmission starting from the first device.\footnote{When an AP coordinates devices to generate new status packets, we consider it the beginning of a new transmission round. In TDMA-NR, the AP may not receive all the status packets in a transmission round (due to transmission failure).} Therefore, the number of time slots in a round is not always the same. This varies depending on the result of packet decoding.

\subsubsection{Instantaneous AoC of TDMA-NR}
We denote the number of time slots in the $j$-th round by $d_j$. The $j$-th round starts transmission at $t_{j-1}$ and ends at $t_j$. Since the total number of devices in the system is $N$, the minimum number of time slots in a successful round is also $N$.  If the status packets of all devices are successfully decoded, the instantaneous AoC drops to $N$. 

\begin{figure}[t]
    \centering
    \includegraphics[width=3.5in]{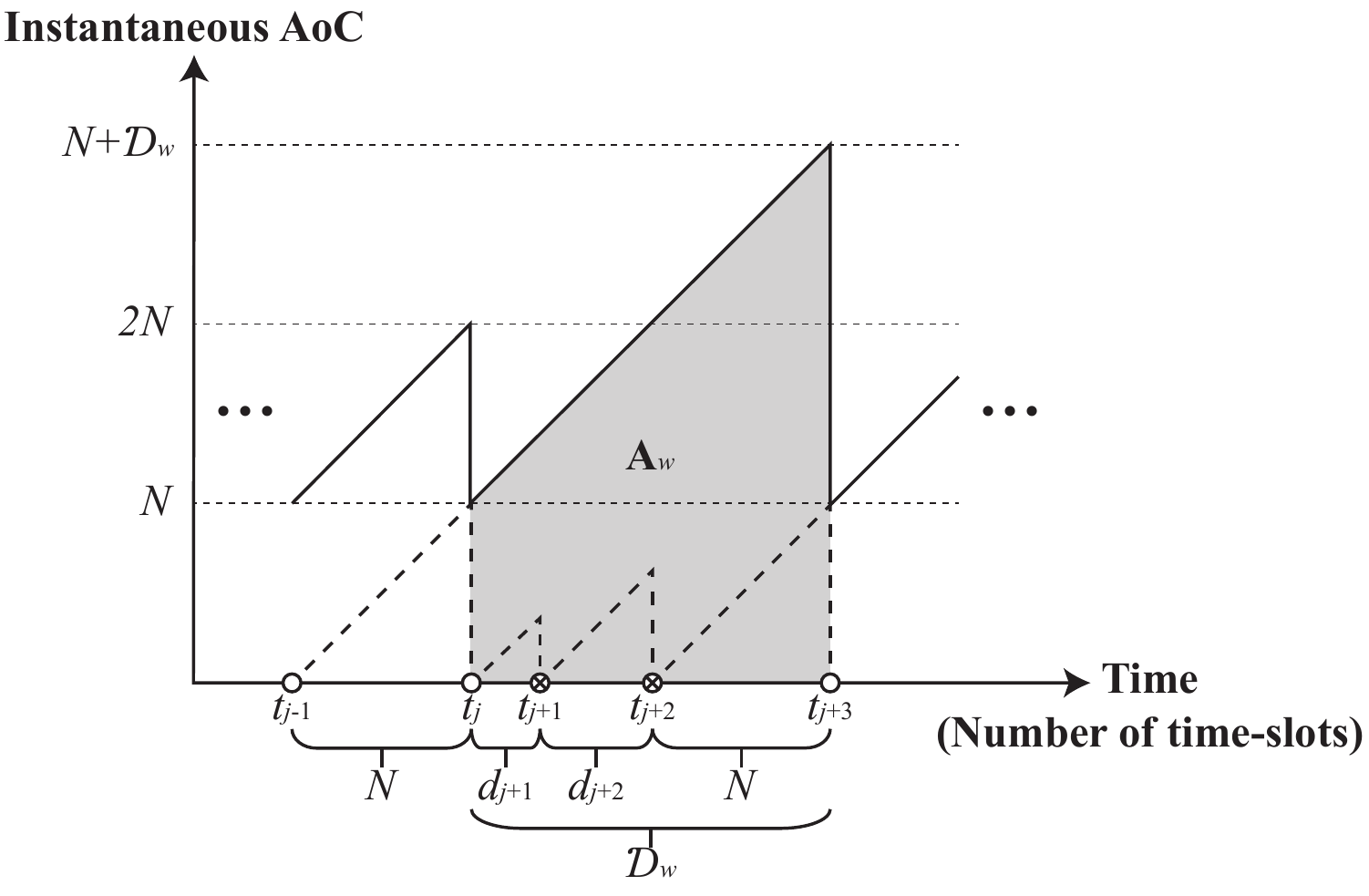}
    \caption{An example of the instantaneous AoC, $\Delta_{TD-NR} (t)$, of an $N$-device system using the TDMA-NR scheme.}
    \label{fig:tdma_nr_aoc}
\end{figure}

Fig. \ref{fig:tdma_nr_aoc} shows an example of the instantaneous AoC of an $N$-device system adopting the TDMA-NR scheme. At $t_{j-1}$, $t_{j}$, and $t_{j+3}$, the instantaneous AoC of the system drops to $N$ as the AP successfully decodes the packets from all devices. At $t_{j+1}$ and $t_{j+2}$, the packets from the devices fail to be decoded, so $\Delta (t)$ continues to increase linearly with time. Suppose the failed devices are device $i'$ in round $j+1$ and device $i''$ in round $j+2$, we can also know that ${d_{j+1}} = i'$ and ${d_{j+2}} = i''$.

\subsubsection{Average AoC of TDMA-NR}
Suppose that the $w$-th successful round is round $j$, ending at $t_j$, and the $(w+1)$-th successful round is round $j+\alpha$, ending at $t_{j+\alpha}$. The number of time slots between $t_j$ and $t_{j+\alpha}$ is ${{\cal D}_w}$, i.e., $t_{j+\alpha} - t_j = {{\cal D}_w}$ (in units of time slots). To compute the average AoC of the TDMA-NR system, let us consider the area under the curve between two consecutive successful updates in Fig. \ref{fig:tdma_nr_aoc}. For example, area ${\bf A}_w$ between $t_j$ and $t_{j+ \alpha}$ can be expressed as
\begin{equation}
    {\bf A}_w = \int_{{t_j}}^{{t_{j + \alpha }} } {\Delta_{TD-NR} (t)\ dt}  = N{\cal D}_w + \frac{{{{{\cal D}_w}^2}}}{2}.
\end{equation}
After obtaining area ${\bf A}_w$, the average AoC is derived as
\begin{align} \label{eqa:tdma_nr_avg_aoc}
{{\bar \Delta }_{TD-NR}} &= \mathop {\lim }\limits_{T \to \infty } \frac{1}{T}\int_0^T {\Delta_{TD-NR} (t)\ dt}  \notag \\
 &= \mathop {\lim }\limits_{W \to \infty } \frac{{\sum\nolimits_{w = 1}^W {{{\bf A}_w}} }}{{\sum\nolimits_{w = 1}^W {{{\cal D}_w}} }} \notag  \\
 &= \mathop {\lim }\limits_{W \to \infty } \frac{{\sum\nolimits_{w = 1}^W {N{{\cal D}_w} + \frac{{{{\cal D}_w}^2}}{2}} }}{{\sum\nolimits_{w = 1}^W {{{\cal D}_w}} }} \notag  \\
 &=  {N + \frac{{E[{{\cal D}^2}]}}{{2E[{\cal D}]}}} 
\end{align}
where $E[{\cal D}]$ and $E[{\cal D}^2]$ are the expected values of ${\cal D}_w$ and ${{\cal D}_w}^2$, respectively. According to \eqref{eqa:tdma_nr_avg_aoc}, we need to derive $E[{\cal D}]$ and $E[{{\cal D}^2}]$ to obtain the average AoC of the TDMA-NR system. 

\subsubsection{Markov chain analysis of TDMA-NR}
\begin{figure}[t]
    \centering
    \includegraphics[width=3.5in]{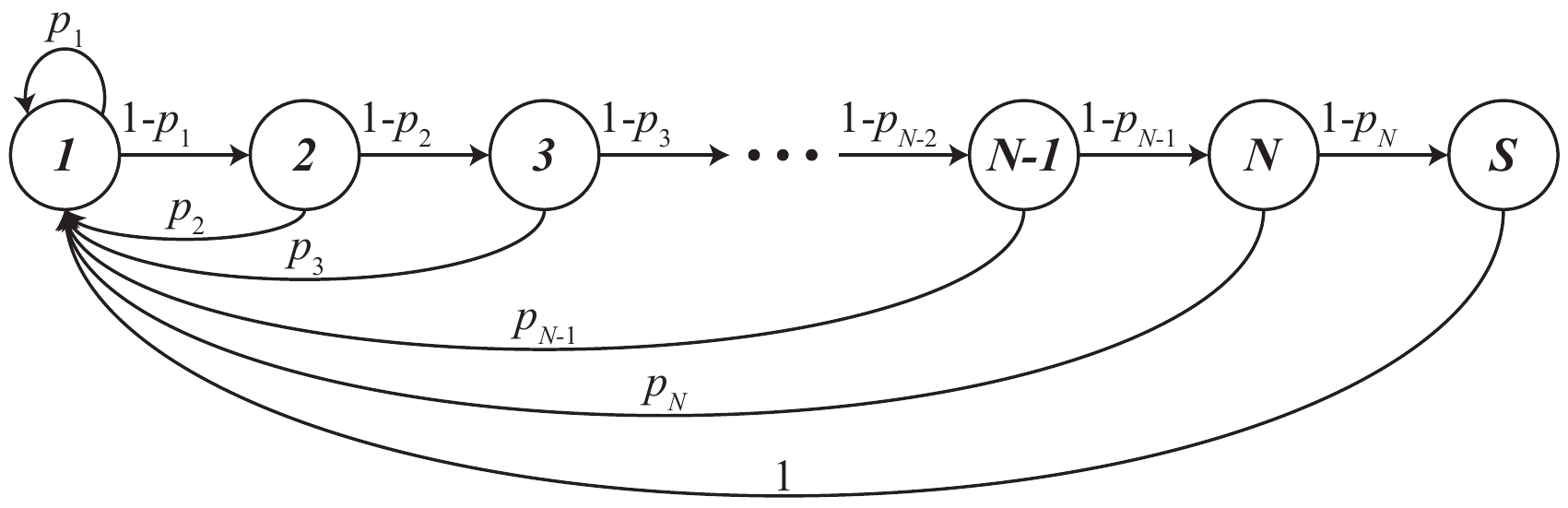}
    \caption{The state transition diagram of TDMA-NR.}
    \label{fig:tdma_no_ret}
\end{figure}

Since which device transmits in the current time slot depends only on the transmission result in the previous slot, we use a Markov chain to model the transmission states in a TDMA-NR system. Fig. \ref{fig:tdma_no_ret} shows the state transition diagram of TDMA-NR. State $i$ in the diagram represents the transmission of device $i$ in time slot $i$ of a transmission round. In state $i$, device $i$ successfully transmits a packet with probability $1-p_i$, where ${p_i} \in [0,1]$ represents the packet error rate (PER) of device $i$. If the transmission of device $i$ is successful, device $i+1$ will transmit in the next time slot. Otherwise, the AP will ask all the devices to generate new status packets and prepare for a new transmission round. If all status packets are successfully decoded, the system will reach state $S$. Then, the instantaneous AoC will drop to $N$, and the AP will coordinate a new transmission round.

\paragraph{Computation of $E[{\cal D}]$}
Since $\cal D$ is the number of time slots between two consecutive successful transmission rounds, $E[{\cal D}]$ is equivalent to the mean hitting time of state $S$ starting from the state $1$. To describe the derivation of $E[{\cal D}]$, we use ${T_{i,S}}$ to denote the mean hitting time of state $S$ starting from state $i$. In other words, $E[{\cal D}]={T_{1,S}}$ in the TDMA-NR system.

Let $X_0$ and $X_1$ represent the current and the next state, respectively, and ${t_0}$ represent the time to reach state $S$ from the current state. The mean hitting time of state $S$ starting from state $1$ is
\begin{align}
{T_{1,S}} &= E\left[ {{t_0}|{X_0} = 1} \right] \notag \\
 &= 1 + \sum\limits_{k \in \{ 1,2\} }^{} {E\left[ {{t_0}|{X_1} = k} \right]P\left( {{X_1} = k|{X_0} = 1} \right)} \notag  \\
 &= 1 + {p_1}{T_{1,S}} + (1 - {p_1}){T_{2,S}}.
\end{align}
Furthermore, we have
\begin{align}
{T_{2,S}} &= E\left[ {{t_0}|{X_0} = 2} \right] \notag  \\
 &= 1 + \sum\limits_{k \in \{ 1,3\} }^{} {E\left[ {{t_0}|{X_1} = k} \right]P\left( {{X_1} = k|{X_0} = 2} \right)} \notag  \\
 &= 1 + {p_2}{T_{1,S}} + (1 - {p_2}){T_{3,S}}.
\end{align} In general, for state $i' \left( {1 \le i' < N} \right)$, we have
\begin{align}
{T_{i',S}} &= E\left[ {{t_0}|{X_0} = i'} \right] \notag  \\
 &= 1 + \sum\limits_{k \in \{ 1,i' + 1\} }^{} {E\left[ {{t_0}|{X_1} = k} \right]P\left( {{X_1} = k|{X_0} = i'} \right)} \notag  \\
 &= 1 + {p_{i'}}{T_{1,S}} + (1 - {p_{i'}}){T_{i' + 1,S}}.
\end{align}
For ${T_{N,S}}$, we have
\begin{equation} \label{eqa:tdma_nr_t_ns}
    {T_{N,S}} = 1 + {p_N}{T_{1,S}} + (1 - {p_N}){T_{S,S}} = 1 + {p_N}{T_{1,S}}.
\end{equation}
The second equality in \eqref{eqa:tdma_nr_t_ns} holds because state $S$ is the destination state and ${T_{S,S}} = 0$.

Finally, we can find $E[{\cal D}]$, i.e., $T_{1,S}$, by solving the matrix equation shown in \eqref{eqa:t_1,1_mat_eqa}, which can be written as ${\bf{MS}} = {\bf{e}}$. We need to calculate the inverse of $\bf{M}$ and multiply it by $\bf{e}$ to get ${T_{1,S}}$ in $\bf{S}$, i.e., ${\bf{S}} = {{\bf{M}}^{ - 1}}{\bf{e}}$.

\begin{figure*}[!t]
\small
\footnotesize
\begin{align}
\underbrace {\left[ {
\begin{array}{*{20}{c}}
{1 - {p_1}}&{ - 1 + {p_1}}&0& 0 & \cdots &0&0\\
{ - {p_2}}&1&{ - 1 + {p_2}}& 0 & \cdots &0&0\\
{ - {p_3}}&0&1& { - 1 + {p_3}} & \cdots &0&0\\
 \vdots & \vdots & \vdots & \vdots & \ddots & \vdots & \vdots \\
{ - {p_{N - 2}}}&0&0&  0 & \cdots &{ - 1 + {p_{N - 2}}}&0\\
{ - {p_{N - 1}}}&0&0&  0 & \cdots &1&{ - 1 + {p_{N - 1}}}\\
{ - {p_N}}&0&0&  0 & \cdots &0&1
\end{array}
} \right]}_{\bf{M}}\underbrace {\left[ {\begin{array}{*{20}{c}}
{{T_{1,S}}}\\
{{T_{2,S}}}\\
{{T_{3,S}}}\\
 \vdots \\
{{T_{N - 1,S}}}\\
{{T_{N,S}}}
\end{array}} \right]}_{\bf{S}} = \underbrace {\left[ {\begin{array}{*{20}{c}}
{1}\\
{1}\\
{1}\\
 \vdots \\
{1}\\
{1}
\end{array}} \right]}_{\bf{e}}.
\label{eqa:t_1,1_mat_eqa}
\end{align}
\hrulefill
\end{figure*}

\begin{figure*}[!t]
\small
\footnotesize
\setcounter{equation}{15}
\begin{align}
\underbrace {\left[ {
\begin{array}{*{20}{c}}
{1 - {p_1}}&{ - 1 + {p_1}}&0& 0 & \cdots &0&0\\
{ - {p_2}}&1&{ - 1 + {p_2}}& 0 & \cdots &0&0\\
{ - {p_3}}&0&1& { - 1 + {p_3}} & \cdots &0&0\\
 \vdots & \vdots & \vdots & \vdots & \ddots & \vdots & \vdots \\
{ - {p_{N - 2}}}&0&0&  0 & \cdots &{ - 1 + {p_{N - 2}}}&0\\
{ - {p_{N - 1}}}&0&0&  0 & \cdots &1&{ - 1 + {p_{N - 1}}}\\
{ - {p_N}}&0&0&  0 & \cdots &0&1
\end{array}
} \right]}_{\bf{M}}\underbrace {\left[ {
\begin{array}{*{20}{c}}
{T_{1,S}^2}\\
\begin{array}{l}
T_{2,S}^2\\
T_{3,S}^2
\end{array}\\
 \vdots \\
{T_{N - 1,S}^2}\\
{T_{N,S}^2}
\end{array}
} \right]}_{{\bf{S'}}} = \underbrace {\left[ {\begin{array}{*{20}{c}}
\begin{array}{c}
1 + 2{p_1}{T_{1,S}} + 2(1 - {p_1}){T_{2,S}}\\
1 + 2{p_2}{T_{1,S}} + 2(1 - {p_2}){T_{3,S}}\\
1 + 2{p_3}{T_{1,S}} + 2(1 - {p_3}){T_{4,S}}
\end{array}\\
 \vdots \\
\begin{array}{c}
{1 + 2{p_{N - 2}}{T_{1,S}} + 2(1 - {p_{N - 2}}){T_{N - 1,S}}}\\
1 + 2{p_{N - 1}}{T_{1,S}} + 2(1 - {p_{N - 1}}){T_{N,S}}\\
1 + 2{p_N}{T_{1,S}}
\end{array}
\end{array}} \right]}_{\bf{N}}.
\label{eqa:t^2_1,1_mat_eqa}
\end{align}
\hrulefill
\end{figure*}
\setcounter{equation}{9}

\paragraph{Computation of $E[{\cal D}^2]$}
We denote the mean of the squared hitting time of state $S$ starting from state $i$ by $T_{i,S}^2$, i.e., $E[{{\cal D}^2}] = T_{1,S}^2$. First, we express $T_{1,S}^2$ as
\begin{align}   \label{eqa:t_1s_sqr}
T_{1,S}^2 &= E\left[ {t_0^2|{X_0} = 1} \right] \notag \\
 &= \sum\limits_{k \in \{ 1,2\} } {E\left[ {{{(1 + {t_0})}^2}|{X_1} = k} \right]P\left( {{X_1} = k|{X_0} = 1} \right)} \notag \\
 &= \sum\limits_{k \in \{ 1,2\} } {E\left[ {1 + 2{t_0} + t_0^2|{X_1} = k} \right]P\left( {{X_1} = k|{X_0} = 1} \right)} \notag \\
 &\begin{aligned}
 = 1 &+ 2\sum\limits_{k \in \{ 1,2\} } {E\left[ {{t_0}|{X_1} = k} \right]P\left( {{X_1} = k|{X_0} = 1} \right)} \\ 
 &+ \sum\limits_{k \in \{ 1,2\} } {E\left[ {t_0^2|{X_1} = k} \right]P\left( {{X_1} = k|{X_0} = 1} \right)}
 \end{aligned} \notag \\
 &\begin{aligned}
 = 1 &+ 2\left( {{p_1}{T_{1,S}} + (1 - {p_1}){T_{2,S}}} \right) \\
 &+ \left( {{p_1}T_{1,S}^2 + (1 - {p_1})T_{2,S}^2} \right).
 \end{aligned}
\end{align}
According to \eqref{eqa:t_1s_sqr}, we have
\begin{equation}
\label{eqa:MS'N_1}
    (1 - {p_1})T_{1,S}^2 - (1 - {p_1})T_{2,S}^2 = 1 + 2{p_1}{T_{1,S}} + 2(1 - {p_1}){T_{2,S}}.
\end{equation}
In general, for state $i \left( {1 \le i \le N} \right)$, $T_{i,S}^2$ can be computed by
\begin{align} \label{eqa:tdma_nr_t_is_sqr}
T_{i,S}^2 &= E\left[ {t_0^2|{X_0} = i} \right] \notag \\
 &= \sum\limits_{k \in \{ 1,i + 1\} } {E\left[ {{{(1 + {t_0})}^2}|{X_1} = k} \right]P\left( {{X_1} = k|{X_0} = i} \right)} \notag \\
 &\begin{aligned}
 = 1 &+ 2\sum\limits_{k \in \{ 1,i + 1\} } {E\left[ {{t_0}|{X_1} = k} \right]P\left( {{X_1} = k|{X_0} = i} \right)} \\
 &+ \sum\limits_{k \in \{ 1,i + 1\} } {E\left[ {t_0^2|{X_1} = k} \right]P\left( {{X_1} = k|{X_0} = i} \right)}
 \end{aligned} \notag \\
 &\begin{aligned}
 = 1 &+ 2\left( {{p_i}{T_{1,S}} + (1 - {p_i}){T_{i + 1,S}}} \right) \\
 &+ \left( {{p_i}T_{1,S}^2 + (1 - {p_i})T_{i + 1,S}^2} \right).
 \end{aligned}
\end{align}
Equation \eqref{eqa:tdma_nr_t_is_sqr} can be simplified as
\begin{align}
\label{eqa:MS'N_2}
    - {p_i}T_{1,S}^2 + T_{i,S}^2 &- (1 - {p_i})T_{i + 1,S}^2 \notag \\
    &= 1 + 2{p_i}{T_{1,S}} + 2(1 - {p_i}){T_{i + 1,S}}.
\end{align}
Specifically for $i=N$, we have
\begin{align}    \label{eqa:general_t_s,s_eqa}
    - {p_N}T_{1,S}^2 + T_{N,S}^2 &- (1 - {p_N})T_{S,S}^2 \notag \\
    &= 1 + 2{p_N}{T_{1,S}} + 2(1 - {p_N}){T_{S,S}}.
\end{align}
As ${T_{S,S}} = 0$ and ${T^2_{S,S}} = 0$,  \eqref{eqa:general_t_s,s_eqa} can be further simplified as
\begin{equation}
\label{eqa:MS'N_3}
    - {p_N}T_{1,S}^2 + T_{N,S}^2 = 1 + 2{p_N}{T_{1,S}}.
\end{equation}

Using \eqref{eqa:MS'N_1}, \eqref{eqa:MS'N_2}, and \eqref{eqa:MS'N_3}, we can form the matrices in \eqref{eqa:t^2_1,1_mat_eqa}. Equation \eqref{eqa:t^2_1,1_mat_eqa} can be written as $\bf{M} \bf{S'} = \bf{N}$, where the first matrix $\bf{M}$ on the LHS of \eqref{eqa:t^2_1,1_mat_eqa} is the same as in \eqref{eqa:t_1,1_mat_eqa}. We can get $T_{1,S}^2$ in ${\bf{S'}}$ by computing ${\bf{S'}} = {{\bf{M}}^{ - 1}}{\bf{N}}$.

Once $E[{\cal D}]$ and $E[{{\cal D}^2}]$ (i.e., $T_{1,S}^{}$ and $T_{1,S}^{2}$) are found, the average AoC can be computed by putting $E[{\cal D}]$ and $E[{{\cal D}^2}]$ into \eqref{eqa:tdma_nr_avg_aoc}. If we take into account the duration of a TDMA time slot, ${T^{TD}}$, we can also express the average AoC of TDMA-NR in absolute time as
\setcounter{equation}{16}
\begin{equation}
\label{TDMA-NR_AoC}
    {\bar \Delta _{TD-NR}} = {T^{TD}} \left( {N + \frac{{T_{1,S}^2}}{{2{T_{1,S}}}}} \right).
\end{equation}

\subsection{TDMA-R}
A major problem with TDMA-NR is that if any device fails to transmit a packet, the entire system needs to start a new transmission round. This is too costly because all other previously transmitted packets are wasted, for example, if all but the last device successfully transmit. To avoid wasting the successful transmissions before the failed ones, we now consider a \textbf{TDMA} scheme with \textbf{R}e-transmission. We refer to this scheme as \textbf{\textit{TDMA-R}}.

On the one hand, TDMA-R may have a shorter interval between consecutive successful updates at the AP, as only the failed update is retransmitted while the successfully transmitted packets are retained. On the other hand, when a device fails to transmit in TDMA-NR, all devices have the opportunity to sample and generate new status packets, which reduces the instantaneous AoC when the AP successfully receives all the packets. It remains an open question which can achieve better average AoC performance, TDMA-NR or TDMA-R. Therefore, we analyze the TDMA-R scheme in the following. After that, we will compare TDMA-NR and TDMA-R comprehensively in Section~\ref{Performance Evaluation}.

In TDMA-R, when the AP fails to decode the packet transmitted by device $i$, the AP uses an ACK to inform all devices of the failure and asks device $i$ to retransmit the same packet in the next time slot. Device $i+1$ will not send its status packet until the AP successfully decodes device $i$'s packet.

There is a special case in TDMA-R: if the AP fails to decode the packet from device $1$, the AP will request all devices to generate new status packets. Intuitively, in this case, sending a new packet from device $1$ instead of retransmitting the old packet always has a lower instantaneous AoC when the AP successfully receives packets from all devices.

\subsubsection{Instantaneous AoC of TDMA-R}
\begin{figure}[t]
    \centering
    \includegraphics[width=3.5in]{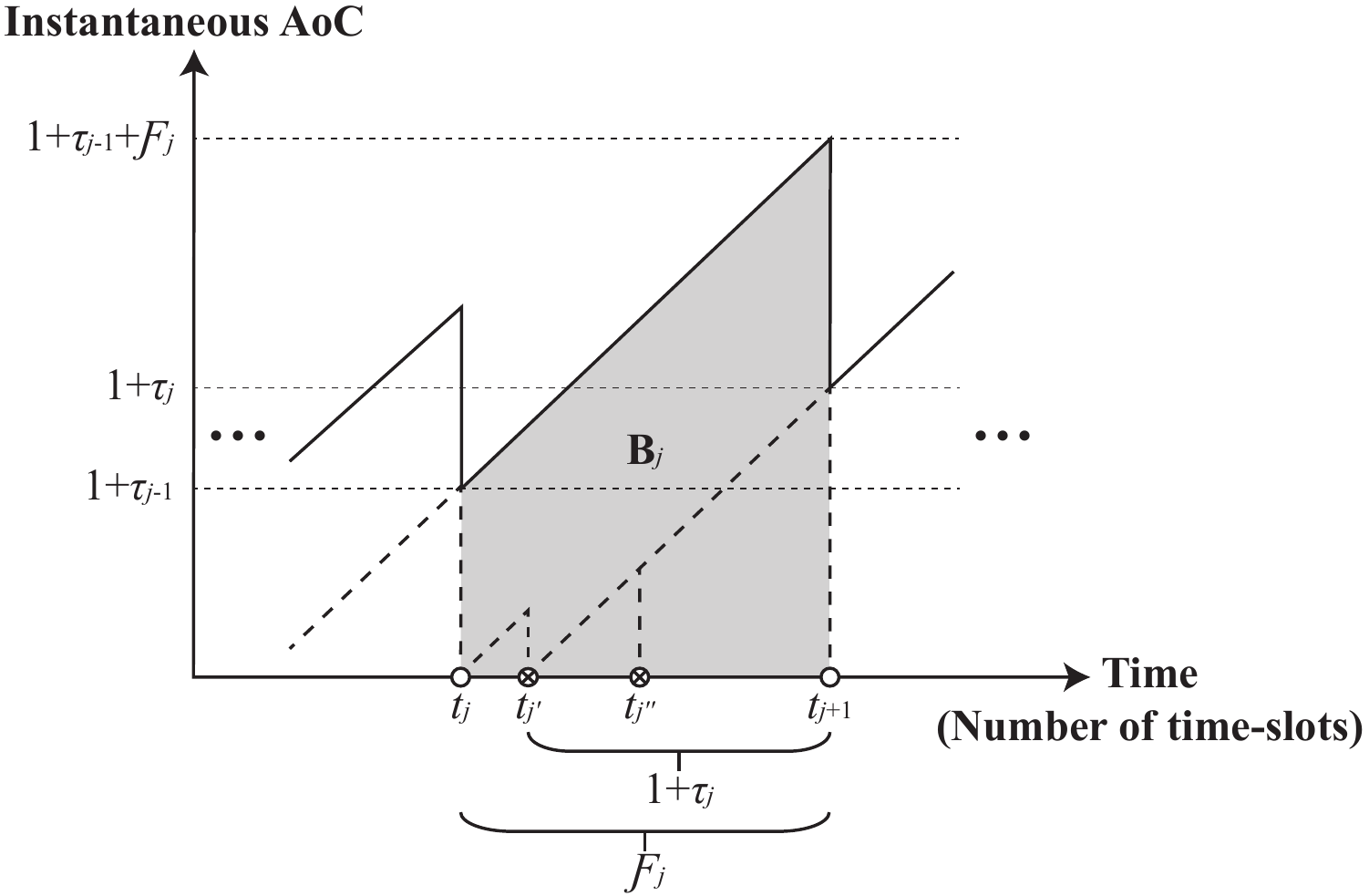}
    \caption{An example of the instantaneous AoC, $\Delta_{TD-R} (t)$, of an $N$-device TDMA-R system.}
    \label{fig:tdma_with_ret}
\end{figure}
We use $t_j$ to denote the start time of round $j$, i.e., $t_j$ is also the end time of round $j-1$, and ${{\cal F}_j}$ to denote the number of time slots in round $j$. Thus, we have ${\cal F}_j = t_{j+1} - t_j$. TDMA-R starts a new transmission round only when the AP successfully decodes the packets from all devices in the previous round.

Recall that the time between generating a new packet by device $1$ and its successful transmission is always one time slot. Suppose that in round $j$, the number of time slots elapsed from the generation of the first packet by device $2$ to the end of the round is ${\tau _{j}}$. In this case, the number of time slots from the generation of a successfully transmitted packet by device $1$ to the end of round $j$ is $1 + {\tau _{j}}$.

Fig. \ref{fig:tdma_with_ret} shows an example of the instantaneous AoC of an $N$-device TDMA-R system. In this example, after the AP has decoded all packets in round $j-1$, the instantaneous AoC is reset to $1 + {\tau _{{j-1}}}$ at $t_j$. After that, the system starts a new transmission round (i.e., the $j$-th round). At ${t_{j'}}$, the status packet of device $1$ fails to be decoded. Then the AP uses an ACK to let all devices generate new status packets for transmission. After several slots of transmission, a device (not device $1$) fails to transmit at ${t_{j''}}$. At this time, the AP broadcasts an ACK and allows the failed device to retransmit the same packet again. Finally, at ${t_{j + 1}}$, the instantaneous AoC drops to $1 + {\tau _{j}}$, because the AP has successfully decoded the packets of all devices in round $j$. Notice that $1 + {\tau _{j}}$ is not a fixed value for each round.

\subsubsection{Average AoC of TDMA-R}
Similar to TDMA-NR, we compute the average AoC by first considering the area under the curve for the $j$-th round of transmission (see area ${{\bf B}_j}$ in Fig. \ref{fig:tdma_with_ret}). The calculation of area ${{\bf B}_j}$ can be written as
\begin{equation}
    {{\bf B}_j} = \int_{t_j}^{{t_{j + 1}}} {\Delta_{TD-R} (t)\ dt}  = {{\cal F}_j} + {{\cal F}_j}{\tau _{j - 1}} + \frac{{{{\cal F}_j}^2}}{2}.
\end{equation}
Then, the average AoC of the TDMA-R system can be computed by
\begin{align}   \label{eqa:tdma_r_aoc_comp}{{\bar \Delta }_{TD-R}} &= \mathop {\lim }\limits_{T \to \infty } \frac{1}{T}\int_0^T {\Delta_{TD-R} (t)\ dt} \notag 
 \\ &= \mathop {\lim }\limits_{J \to \infty } \frac{{\sum\nolimits_{j = 1}^J {{{\bf B}_j}} }}{{\sum\nolimits_{j = 1}^J {{{\cal F}_j}} }} \notag \\
&= \mathop {\lim }\limits_{J \to \infty } \frac{{\sum\nolimits_{j = 1}^J {{{\cal F}_j} + {{\cal F}_j} {\tau}_{j-1} +\frac{{{{\cal F}_j}^2}}{2}} }}{{\sum\nolimits_{j = 1}^J {{{\cal F}_j}} }} \notag \\
&= {1 + E[{\tau}] + \frac{{E[{{\cal F}^2}]}}{{2{E[{\cal F}]}}}}
\end{align}
where $E[\cdot]$ denotes the expected value of the random variables. Note that ${\cal F}_j$ and $\tau_{j-1}$ depend on the current and previous transmission rounds, respectively. Thus, they are independent random variables.

\subsubsection{Markov chain analysis of TDMA-R}
\begin{figure}[t]
    \centering
    \includegraphics[width=3.5in]{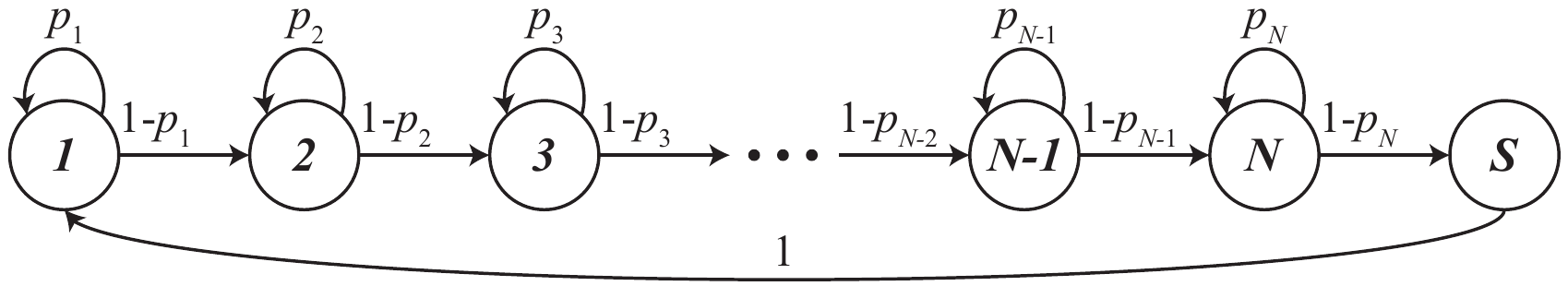}
    \caption{The state transition diagram of TDMA-R.}
    \label{fig:tdma_with_ret_diagram}
\end{figure}

In TDMA-R, the transmission in the current time slot depends only on the transmission result in the previous time slot. Therefore, we use a Markov chain to model the state transitions of the TDMA-R system. The state transition diagram is shown in Fig. \ref{fig:tdma_with_ret_diagram}.

In the following, we continue to use ${T_{i,S}}$ to denote the mean hitting time (in time slots) of state $S$ starting from state $i$. We also use $T_{i,S}^2$ to denote the mean of the squared hitting time of state $S$ starting from state $i$. Obviously, ${{E[{\cal F}]}}$ is equivalent to ${T_{1,S}}$ and ${E[{\tau}]}$ is equivalent to ${T_{2,S}}$.

\paragraph{Computation of ${E[{\cal F}]}$ $\left({T_{1,S}} \right)$ and ${E[{\tau}]}$ $\left({T_{2,S}} \right)$}
${T_{i,S}}$ can be considered as the sum of the means of the geometric random variables with parameters $1-p_k$, where $i \le k \le N$, since the latter device in a transmission round can start transmission only after the packet of the former device has been successfully decoded. In other words, we can compute ${T_{i,S}}$ by
\begin{equation}    \label{eqa:tdma_t_i,s}
    {T_{i,S}} = \sum\limits_{k = i}^N {\frac{1}{{1 - {p_k}}}}
\end{equation}
where $1 \le i \le N$. In addition, we have ${T_{S,S}} = 0$ as state $S$ is the destination state. Thus, ${E[{\cal F}]}$ $\left({T_{1,S}} \right)$ and ${E[{\tau}]}$ $\left({T_{2,S}} \right)$ can be found by
\begin{align}
    {E[{\cal F}]} &= {T_{1,S}} = \sum\limits_{k = 1}^N {\frac{1}{{1 - {p_k}}}}, \\        {E[{\tau}]} &= {T_{2,S}} = \sum\limits_{k = 2}^N {\frac{1}{{1 - {p_k}}}}.
\end{align}

\paragraph{Computation of ${E[{\cal F}^2]}$ $\left({T_{1,S}^2} \right)$}
We now show the computation of $E[{{\cal F}^2}]$, i.e., ${T_{1,S}^2}$. For $1 \le i \le N$, we have
\begin{align}
\label{E[F^2]}
T_{i,S}^2 &= E\left[ {t_0^2|{X_0} = i} \right] \notag \\
 &= \sum\limits_{k \in \{ i,i + 1\} } {E\left[ {{{(1 + {t_0})}^2}|{X_1} = k} \right]P\left( {{X_1} = k|{X_0} = i} \right)} \notag \\
&\begin{aligned}
= 1 &+ 2\sum\limits_{k \in \{ i,i + 1\} } {E\left[ {{t_0}|{X_1} = k} \right]P\left( {{X_1} = k|{X_0} = i} \right)} \\ 
&+ \sum\limits_{k \in \{ i,i + 1\} } {E\left[ {t_0^2|{X_1} = k} \right]P\left( {{X_1} = k|{X_0} = i} \right)} 
\end{aligned} \notag \\
&\begin{aligned}
= 1 &+ 2\left( {{p_i}{T_{i,S}} + (1 - {p_i}){T_{i + 1,S}}} \right) \\
&+ \left( {{p_i}T_{i,S}^2 + (1 - {p_i})T_{i + 1,S}^2} \right).
\end{aligned}
\end{align}
From \eqref{E[F^2]}, we have
\begin{equation}
    T_{i,S}^2 - T_{i + 1,S}^2 = \frac{1}{{1 - {p_i}}} + \frac{{2{p_i}}}{{1 - {p_i}}}{T_{i,S}} + 2{T_{i + 1,S}}.
\end{equation}
In addition, $T_{i,S}^2$ can also be expressed as
\begin{align}   \label{eqa:tdma_r_t^2_i,s}
&\begin{aligned}
T_{i,S}^2 = (T_{i,S}^2 &- T_{i + 1,S}^2) + ({T_{i + 1,S}^2 - T_{i + 2,S}^2}) +  \cdots  \\
&+({T_{N - 1,S}^2 - T_{N,S}^2}) + ( {T_{N,S}^2 - T_{S,S}^2}) + T_{S,S}^2 
\end{aligned} \notag \\
&\begin{aligned}
= &\left( {\frac{1}{{1 - {p_i}}} + \frac{{2{p_i}}}{{1 - {p_i}}}{T_{i,S}} + 2{T_{i + 1,S}}} \right) \\
&+ \left( {\frac{1}{{1 - {p_{i + 1}}}} + \frac{{2{p_{i + 1}}}}{{1 - {p_{i + 1}}}}{T_{i + 1,S}} + 2{T_{i + 2,S}}} \right) \\ 
&+  \cdots \\
&+ \left( {\frac{1}{{1 - {p_{N - 1}}}} + \frac{{2{p_{N - 1}}}}{{1 - {p_{N - 1}}}}{T_{N - 1,S}} + 2{T_{N,S}}} \right) \\
&+ \left( {\frac{1}{{1 - {p_N}}} + \frac{{2{p_N}}}{{1 - {p_N}}}{T_{N,S}}} \right)
\end{aligned} \notag \\
&\begin{aligned}
= &\left( {\frac{1}{{1 - {p_i}}} + \frac{1}{{1 - {p_{i + 1}}}} +  \cdots  + \frac{1}{{1 - {p_{N - 1}}}} + \frac{1}{{1 - {p_N}}}} \right) \\ &+ 
\frac{{2{p_i}}}{{1 - {p_i}}}{T_{i,S}} + \frac{2}{{1 - {p_{i + 1}}}}{T_{i + 1,S}} + \frac{2}{{1 - {p_{i + 2}}}}{T_{i + 2,S}} \\ &+  \cdots  + \frac{2}{{1 - {p_N}}}{T_{N,S}}
\end{aligned} \notag \\
&\begin{aligned}
= {T_{i,S}} &+ \frac{{2{p_i}}}{{1 - {p_i}}}{T_{i,S}} + \frac{2}{{1 - {p_{i + 1}}}}{T_{i + 1,S}} + \frac{2}{{1 - {p_{i + 2}}}}{T_{i + 2,S}} \\ &+  \cdots + \frac{2}{{1 - {p_N}}}{T_{N,S}}
\end{aligned} \notag \\
&\begin{aligned}
= \frac{{1 + {p_i}}}{{1 - {p_i}}}{T_{i,S}} &+ \frac{2}{{1 - {p_{i + 1}}}}{T_{i + 1,S}} + \frac{2}{{1 - {p_{i + 2}}}}{T_{i + 2,S}} \\ &+  \cdots + \frac{2}{{1 - {p_N}}}{T_{N,S}}.
\end{aligned}
\end{align}
Note that ${T_{S,S}} = 0$ and $T_{S,S}^2=0$, because state $S$ is the destination state.

To get $T_{1,S}^2$, we can substitute $i=1$ into \eqref{eqa:tdma_r_t^2_i,s} and obtain
\begin{align}   \label{eqa:tdma_r_t^2_eqa}
T_{1,S}^2 = &\frac{{1 + {p_1}}}{{1 - {p_1}}}{T_{1,S}} + \frac{2}{{1 - {p_2}}}{T_{2,S}} \notag \\ &+  \frac{2}{{1 - {p_3}}}{T_{3,S}} +  \cdots + \frac{2}{{1 - {p_N}}}{T_{N,S}}.
\end{align}
${T_{i,S}}$ for all $i$ can be computed using \eqref{eqa:tdma_t_i,s} and then inserted into \eqref{eqa:tdma_r_t^2_eqa} to obtain $T_{1,S}^2$. Finally, we compute the average AoC of TDMA-R by inserting $T_{1,S}$, $T_{2,S}$, and $T_{1,S}^{2}$ (i.e., $E[{\cal F}]$, $E[{\tau}]$, and $E[{\cal F}^2]$) into \eqref{eqa:tdma_r_aoc_comp}. If we consider the duration of a TDMA time slot, ${T^{TD}}$, the average AoC of TDMA-R in absolute time is
\begin{align}
\label{TDMA-R_AoC}
{\bar \Delta _{TD-R}} = T^{TD} \left( {1 + {T_{2,S}} + \frac{{T_{1,S}^2}}{{2{T_{1,S}}}}} \right).
\end{align}

\section{AoC of FDMA}
\label{AoC of FDMA}
In contrast to TDMA systems, where packets from different devices are transmitted one by one, packets in FDMA systems are transmitted by all devices simultaneously. In the following, we will study the instantaneous and average AoC of FDMA and then compare the performance of TDMA-NR, TDMA-R, and FDMA systems.

\begin{figure}[t]
	\centering
	\includegraphics[width=3.5in]{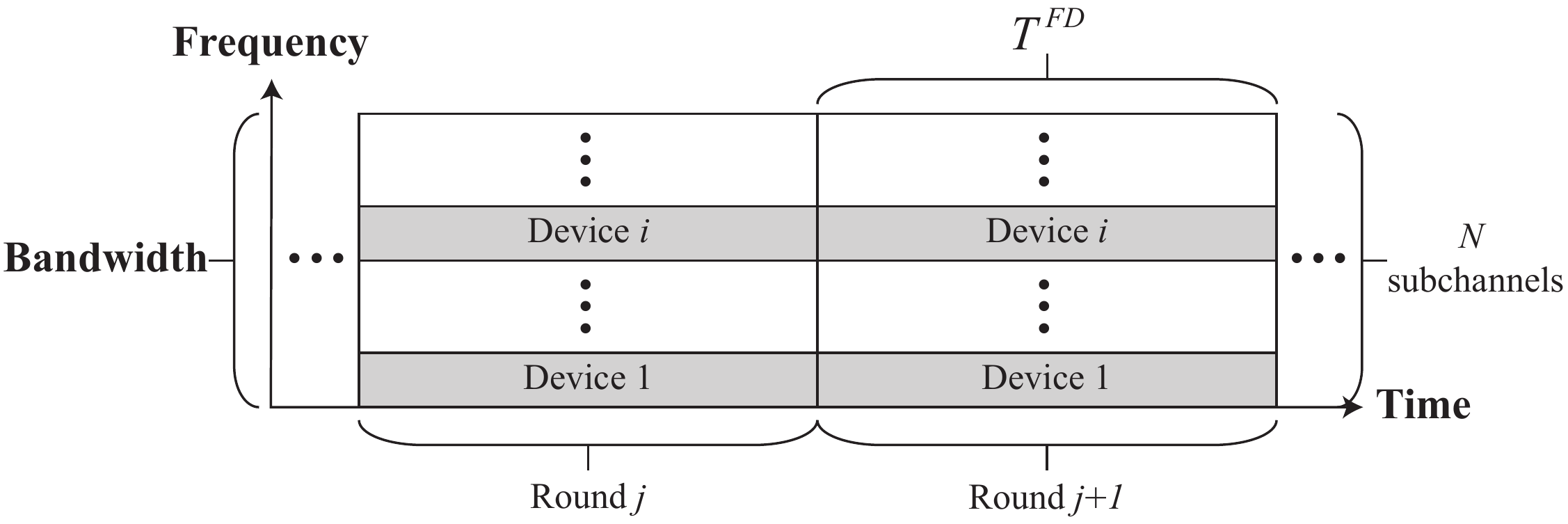}
	\caption{The FDMA scheme with $N$ devices accessing the channel simultaneously.}
	\label{fig:fdma}
\end{figure}

Fig. \ref{fig:fdma} illustrates the FDMA scheme, where the wireless channel is divided into $N$ sub-channels. Since the total bandwidth of the channel is $B$, each sub-channel has a bandwidth of $\frac{B}{N}$. Each IoT device occupies one sub-channel and sends its status packet to the AP at the same time in each transmission round. The device only needs to continuously sample the latest status and generate a new update packet at the beginning of a round.

We assume each round has the same duration ${T^{FD}}$. In transmission round $j$, all devices sample their current status, generate update packets, and send them to the AP simultaneously. The AP then decodes the status packets from all the devices at the end of the round. If all the status packets are decoded successfully, the instantaneous AoC is reset to ${T^{FD}}$.

\subsection{Instantaneous AoC of FDMA}
We use the duration of one round (also one time slot, $T^{FD}$) as a unit in this section, and we will switch the unit of the final result back to time. The probability that a packet from device $i$ is successfully decoded in a transmission round is $(1 - {p_i})$, where $p_i$ denotes the PER and ${p_i} \in [0,1]$. Due to the different channels, different devices have different PERs. Suppose that the $w$-th successful round is round $j$, ending at $t_j$, and the $(w+1)$-th successful round is round $j+\beta_w$, ending at $t_{j+\beta_w}$. In other words, the number of time slots between $t_j$ and $t_{j+\beta_w}$ is $\beta_w$. Then $\beta_w$ is a geometric random variable with a probability mass function of
\begin{equation}
	P\left( {\beta_w = x} \right) = {\left( {1 - \gamma } \right)^{x - 1}}\gamma
\end{equation}
where $\gamma$ is the probability that all the packets are successfully decoded by the AP, i.e., $\gamma  = \prod\limits_{i = 1}^N {\left( {1 - {p_i}} \right)} $.

Fig. \ref{fig:fdma_aoc} shows an example of instantaneous AoC in an $N$-device FDMA system. At $t_{j}$, $t_{j+2}$, and $t_{j+3}$, the instantaneous AoC $\Delta (t)$ is reset to $1$  because packets from all devices are successfully decoded. $\Delta (t)$ continues to increase linearly with time after $t_{j+1}$ because the AP does not successfully decode all the packets in round $j+1$.

\subsection{Average AoC of FDMA}
\begin{figure}[t]
	\centering
	\includegraphics[width=3.5in]{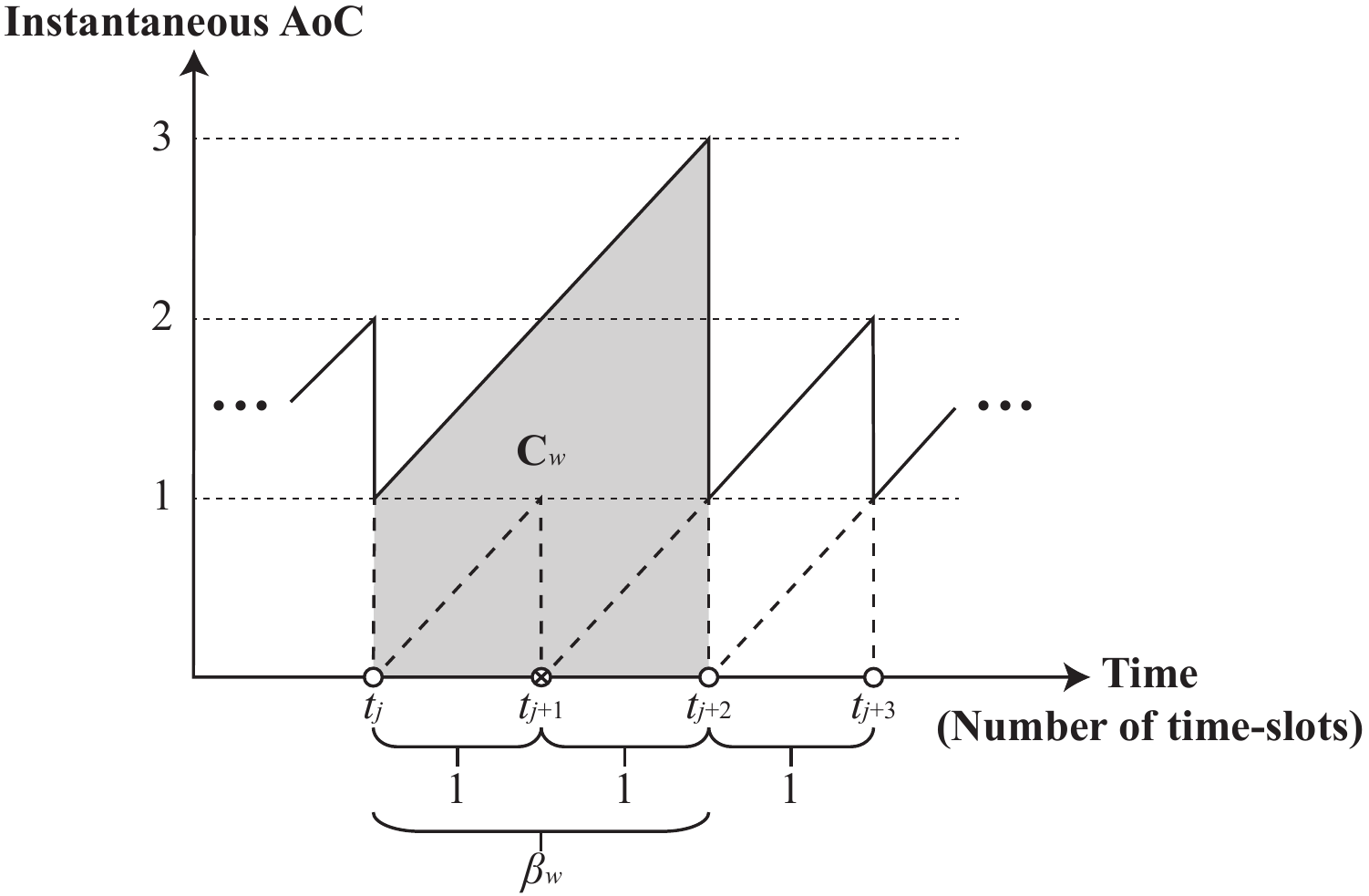}
	\caption{An example of the instantaneous AoC, $\Delta_{FD} (t)$, of an $N$-device FDMA system.}
	\label{fig:fdma_aoc}
\end{figure}

Similar to obtaining the average AoC of TDMA, we can get the average AoC of FDMA by computing the area under the curve between two consecutive successful decoding rounds (e.g., area ${\bf C}_w$ for the $w$-th successful update in Fig. \ref{fig:fdma_aoc}). Area ${\bf C}_w$ can be computed by
\begin{equation}
	{\bf C}_w = \int_{ {t_j}}^{{t_{j + \beta_w}}} {\Delta_{FD} (t)\ dt}  = \beta_w + \frac{{\beta_w}^2}{2}.
\end{equation}
Thus, the average AoC of FDMA, ${\bar \Delta _{FD}}$, is
\begin{align}   \label{eqa:avg_aoc_fdma}
		{{\bar \Delta }_{FD}} &= \mathop {\lim }\limits_{T \to \infty } \frac{1}{T}\int_0^T {\Delta_{FD} (t)\ dt} \notag  \\
		&= \mathop {\lim }\limits_{W \to \infty } \frac{{\sum\nolimits_{w = 1}^W {{{\bf C}_w}} }}{{\sum\nolimits_{w = 1}^W {{\beta _w}} }} \notag  \\
		&=\mathop {\lim }\limits_{W \to \infty } \frac{{\sum\nolimits_{w = 1}^W {\beta_w + \frac{{{{\beta_w}^2}}}{2}} }}{{\sum\nolimits_{w = 1}^W {{\beta _w}} }} \notag  \\
		&= 1 + \frac{{E[{\beta ^2}]}}{{2E[\beta ]}},
\end{align}
where $E[\cdot]$ denotes the expected value of the random variables. Furthermore, for the geometric random variable $\beta_w$, we have
\begin{align} \label{eqa:fdma_expectation}
	E[\beta ] &= \frac{1}{\gamma}, \text{ and} \\
\label{eqa:fdma_sqr_expectation}
	E[{\beta ^2}] &= \frac{{2 - \gamma }}{{{\gamma ^2}}}.
\end{align}
Putting \eqref{eqa:fdma_expectation} and \eqref{eqa:fdma_sqr_expectation} into \eqref{eqa:avg_aoc_fdma} and converting the unit back to time, we have
\begin{equation}
\label{FDMA_AoC}
	{\bar \Delta _{FD}} = T^{FD} \left( {1 + \frac{{2 - \gamma }}{{2\gamma }}} \right).
\end{equation}

\section{Performance Evaluation}
\label{Performance Evaluation}

In this section, we evaluate the AoC performance of TDMA-NR, TDMA-R, and FDMA schemes through USRP SDR experiments with the setup detailed in Section \ref{Experimental Setup}. Section \ref{Theoretical Performance} first depicts the theoretical average AoC of the three multiple access schemes derived in Sections \ref{AoC of TDMA} and \ref{AoC of FDMA}. Subsequently, we validate these results by performance simulations. The theoretical results consider an ideal power-balanced scenario, i.e., each node has the same SNR at the AP. However, the AoC performance of these schemes in practical implementation remains an open question. To address this issue, Section \ref{Experimental Evaluation} examines the practical performance of these schemes in USRP experiments with MAC protocols considered. Moreover, both power-balanced and power-imbalanced cases are considered to analyze the AoC performance of the schemes comprehensively.

\subsection{Experimental Setup}
\label{Experimental Setup}

\begin{figure}[t]
    \centering
    \includegraphics[width=3.5in]{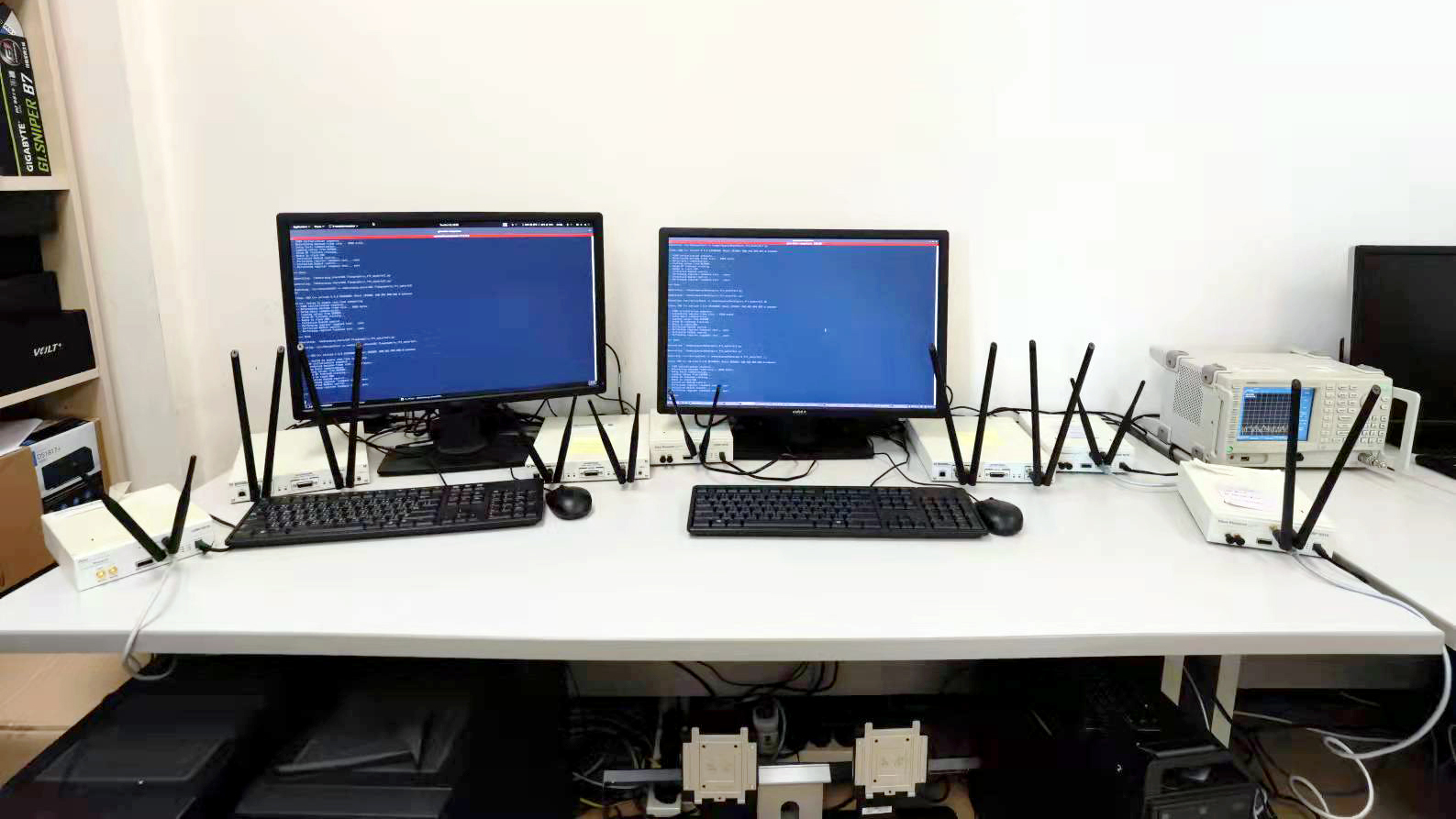}
    \caption{Experimental setup of the indoor information update system.}
    \label{fig:Experiment_AoC}
\end{figure}

We implement the three multiple access schemes on a USRP + GNU Radio platform, as shown in Fig. \ref{fig:Experiment_AoC}. Seven nodes are deployed in an indoor environment, in which each radio node consists of a PC and a USRP connected via an Ethernet cable. One of the radio nodes is selected as the AP, while the rest act as IoT devices. In other words, this configuration constitutes a $6$-user information update system.

We use a mix of USRP X310s and USRP N210s as the radio nodes. We selected a radio node with a USRP X310 as the AP. The USRP hardware driver (UHD) is responsible for the communication between GNU Radio and the USRP hardware. Programs in GNU Radio can use the functions provided by the UHD to control a USRP. In the USRP + GNU Radio SDR platform, the USRP is responsible for (1) transmitting and receiving radio frequency (RF) signals to and from the radio channel and (2) up-converting (down-converting) the baseband (RF) signals to RF (baseband) signals. Meanwhile, the PC is responsible for preparing the transmit baseband signals or processing the received baseband signals. The preparation and processing are done on the GNU Radio platform, an open-source software platform tailored for real-time signal processing on General Purpose Platforms (GPP). All the signal processing, including PSK modulation/demodulation, OFDM modulation/demodulation, channel encoding/decoding, packet detection, and synchronization, is implemented on GNU Radio.

\begin{table}[t]
	\caption{\textcolor{black}{PHY-layer Parameters in the Experiments}}\label{tab:exp}
	\centering
	\def\arraystretch{1}
	{\color{black}\begin{tabular}{l|c}
			\hline
			Center frequency   & $2.418$ GHz               \\ \hline
			Total bandwidth    & $10$ MHz                  \\ \hline
			Size of status packet & $96$ bits \\ \hline
			Modulation         & BPSK                     \\ \hline
			Channel code       & $1/2$ convolutional code \\ \hline
			Length of cyclic prefix (CP) & $16$ samples         \\ \hline
			Guard interval (GI)	& $16$  $\mu s$ \\	\hline
			FFT size (${N_{{\rm{FFT}}}}$)  & $64$			\\ \hline
			Number of subcarriers &  $48$ (TDMA), $8$ (FDMA)		\\ \hline
	\end{tabular}}
\end{table}

In the experiments, we evaluate the average AoC of the two TDMA schemes using the TDMA implementation in \cite{liang_design_2021} and the average AoC of the FDMA scheme using the OFDMA\footnote{OFDMA can be regarded as an FDMA system with tighter gaps between different sub-channels.} implementation introduced in \cite{liang_rofa_2021}.
The physical layer parameters shown in Table~\ref{tab:exp} are used for both TDMA and FDMA schemes. Specifically, we run experiments in a $10$ MHz channel transmitting $96$-bit status packets, a typical packet length for status update systems. BPSK modulation and $1/2$ convolutional code are deployed to ensure transmission reliability.

\subsection{Theoretical Performance}
\label{Theoretical Performance}

\begin{figure}[t]
    \centering
    \includegraphics[width=3.5in]{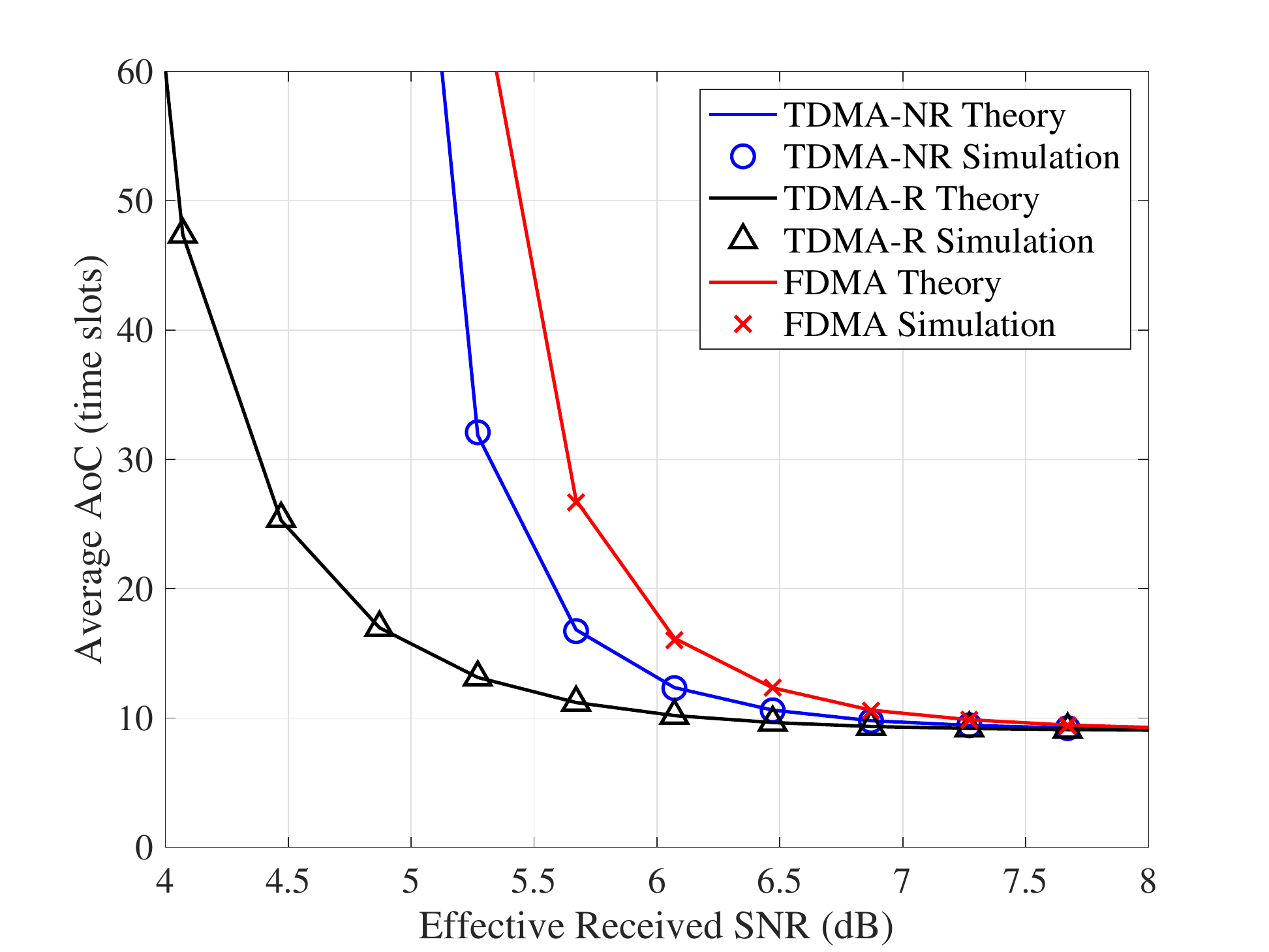}
    \caption{The theoretical average AoC performance of the $6$-user information update system in a power-balanced scenario.}
    \label{fig:aoc_balanced_nomac}
\end{figure}

This subsection presents the theoretical average AoC performance of TDMA-NR, TDMA-R, and FDMA schemes in the power-balanced scenario, as illustrated in Fig. \ref{fig:aoc_balanced_nomac}. We first obtain the packet error rates (PERs) of the TDMA and FDMA transmissions from the decoding results of $10^5$ transmissions for each SNR value in the USRP experiments. Then, we incorporate the PERs into (\ref{TDMA-NR_AoC}), (\ref{TDMA-R_AoC}), and (\ref{FDMA_AoC}) to compute the theoretical average AoC of TDMA-NR, TDMA-R, and FDMA, respectively. Here, we use the number of TDMA time slots as units, and the unit conversion between FDMA and TDMA time slots is given by ${T^{FD}} = N{T^{TD}}$.

To verify the theoretical results, we simulate the evolution of AoC values over time and compute the average AoC. As depicted in Fig.~\ref{fig:aoc_balanced_nomac}, the theoretical results coincide with the simulation results, confirming our theoretical analysis of the average AoC performance. From the results, TDMA-R outperforms both TDMA-NR and FDMA in terms of the average AoC. The lower effective received SNR makes the difference in average AoC among the three schemes more pronounced. The performance gain of TDMA-R comes mainly from avoiding unnecessary retransmissions of successfully delivered packets when a packet fails to be decoded.

\subsection{Experimental Evaluation}
\label{Experimental Evaluation}

In our theoretical analysis and simulation evaluation, we simplify the system model by considering only the transmission time occupied by the status data (i.e., information updates). In addition, the previous subsection considers only the ideal case where each node has the same SNR at the AP. Nevertheless, the impact of the MAC protocol design on the AoC performance and the power-imbalanced case should also be considered, as they are crucial to understanding the information freshness of the cooperative status updates in different multiple access schemes for practical applications. Therefore, this subsection presents an experimental evaluation of the AoC performance in a practical implementation, considering not only the power-imbalanced case but also the transmission time caused by parts other than the status data, such as preamble, ACK, and guard interval (GI). As we will see, TDMA-R may not be the best-performing solution in some cases when these practical factors are taken into account.

\subsubsection{MAC Protocols of Different Schemes}
We adapt the previous general-purpose OFDM systems \cite{liang_design_2021} to our timely status update systems for both TDMA and FDMA. The modifications include (1) designing MAC protocols for different multiple access schemes, and (2) designing short packet formats, including short payloads and reduced preambles to reduce the average AoC, as detailed below.

\paragraph{TDMA-NR and TDMA-R MAC Protocol}
\begin{figure}[t]
    \centering
    \includegraphics[width=3.5in]{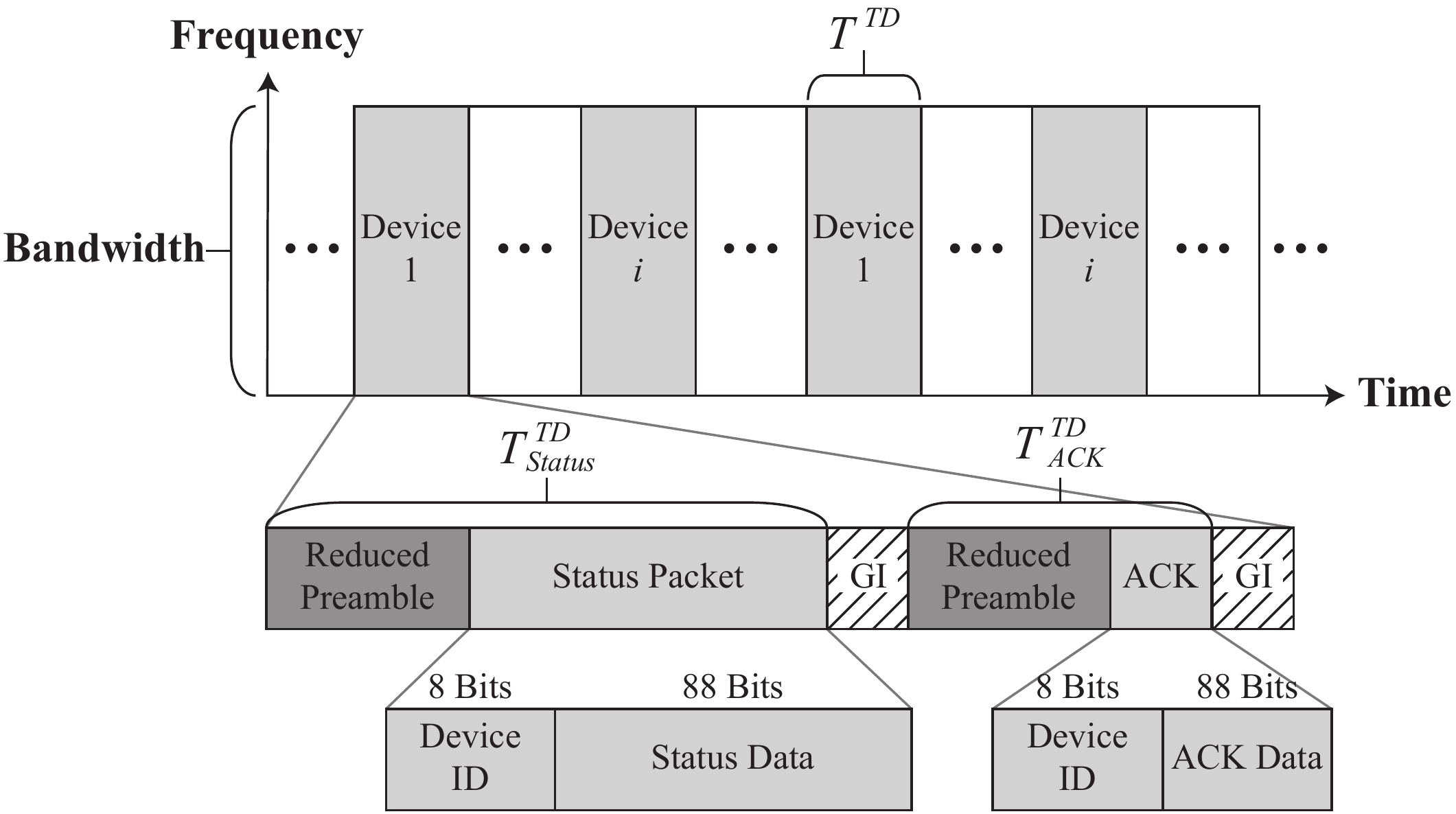}
    \caption{The MAC protocol of the TDMA information update system.}
    \label{fig:tdma_frame}
\end{figure}
The MAC protocol for both TDMA-NR and TDMA-R is shown in Fig. \ref{fig:tdma_frame}. In TDMA-NR and TDMA-R systems, each time slot consists of a status packet transmission period, an ACK period, and two guard intervals (GIs). The status packets have the same block length and duration  $T_{Status}^{TD}$ for all the devices. After a status packet transmission, the AP sends an ACK to the devices during the ACK period. This ACK can either inform the active device about the previous packet reception status or coordinate the devices for a new round of status packet generation and transmission. The duration of an ACK packet is $T_{ACK}^{TD}$. A GI with duration $T_{GI}^{TD}$ is inserted between the status update packet and the ACK packet to ensure sufficient turn-around time for the transmitter and to avoid possible inter-packet interference due to propagation delays.

Now we look at the structure of the status update packet and the ACK shown in Fig. \ref{fig:tdma_frame}. Each update packet or ACK is preceded by a preamble before the payload. The preamble assists the AP in packet synchronization and channel estimation. The status packet payload starts with an $8$-bit device ID field (thus, the total capacity of devices is $2^8 = 256$), followed by $88$ bits of status data. The ACK also starts with an $8$-bit device ID field and a $16$-bit ACK field. The ACK field includes control commands for status packet regeneration and retransmission.

In conventional OFDM systems, Short Training Sequence (STS) is usually used for packet detection. Since the AP in the TDMA system knows the arrival time of each packet, the AP does not need to detect the arrival of a packet before starting the reception process. Instead, the AP can start the packet reception and decoding process by counting the time. Thus, all STSs in the device packets can be eliminated. The elimination of STS can also be applied to ACK packets because the devices are aware of the arrival time of ACK packets. However, we keep the Long Training Sequences (LTSs) because they are used for channel estimation and equalization. We refer to the preamble with STS eliminated as a reduced preamble (RP).

Then, the duration of a status packet with an RP is
\begin{equation}	\label{eqa:td_status_duration}
    {T^{TD}_{Status}} = \left( 160 + \frac{{192}}{{48}} \times 80 \right) \times {10^{ - 4}} = 0.048\text{ ms}.
\end{equation}
Here, $160$ is the number of samples in the RP. The device ID and status data utilize $8$ bits and $88$ bits, respectively, resulting in a total of $96$ bits. These $96$ bits data are subsequently channel-encoded into $192$ coded bits. While a symbol contains $64$ subcarriers, only $48$ subcarriers are allocated for data transmission. Therefore, the number of symbols in a status packet is $(192/48)$. In addition, one symbol comprises a $64$-FFT symbol and a $16$-sample cyclic prefix (CP), amounting to a total of $80$ samples per symbol. Finally, considering a bandwidth of $10$ MHz, the duration of one sample is ${10^{ - 4}}$ ms.

Similarly, the duration of an ACK can be computed by
\begin{equation}
    {T^{TD}_{ACK}} = \left( 160 + \frac{{48}}{{48}} \times 80 \right) \times {10^{ - 4}} = 0.024\text{ ms}.
\end{equation}
Adding the durations of one status packet with an RP, one ACK  with an RP, and two GIs, we can compute the total duration of a TDMA frame as
\begin{equation}
    {T^{TD}} = 0.048 + 0.024 + 0.016 \times 2 = 0.104\text{ ms}.
\end{equation}

\paragraph{FDMA MAC Protocol} 
We demonstrate the MAC protocol of the FDMA system by modifying the implementation of the OFDMA system in \cite{liang_rofa_2021}. Thanks to the phase synchronization among the nodes in the implementation, the interference between adjacent sub-channels is negligible, resulting in the same performance of FDMA. Although there are 
$64$ subcarriers for a symbol, only $48$ subcarriers are used for data transmission. The channel is divided equally into $6$ sub-channels for $6$ users. Each sub-channel occupies $48/6 = 8$ subcarriers. The subcarrier allocation for the $6$ users is shown in Table \ref{tab:fdma_subc_alloc}.

\begin{figure}[t]
    \centering
    \includegraphics[width=3.5in]{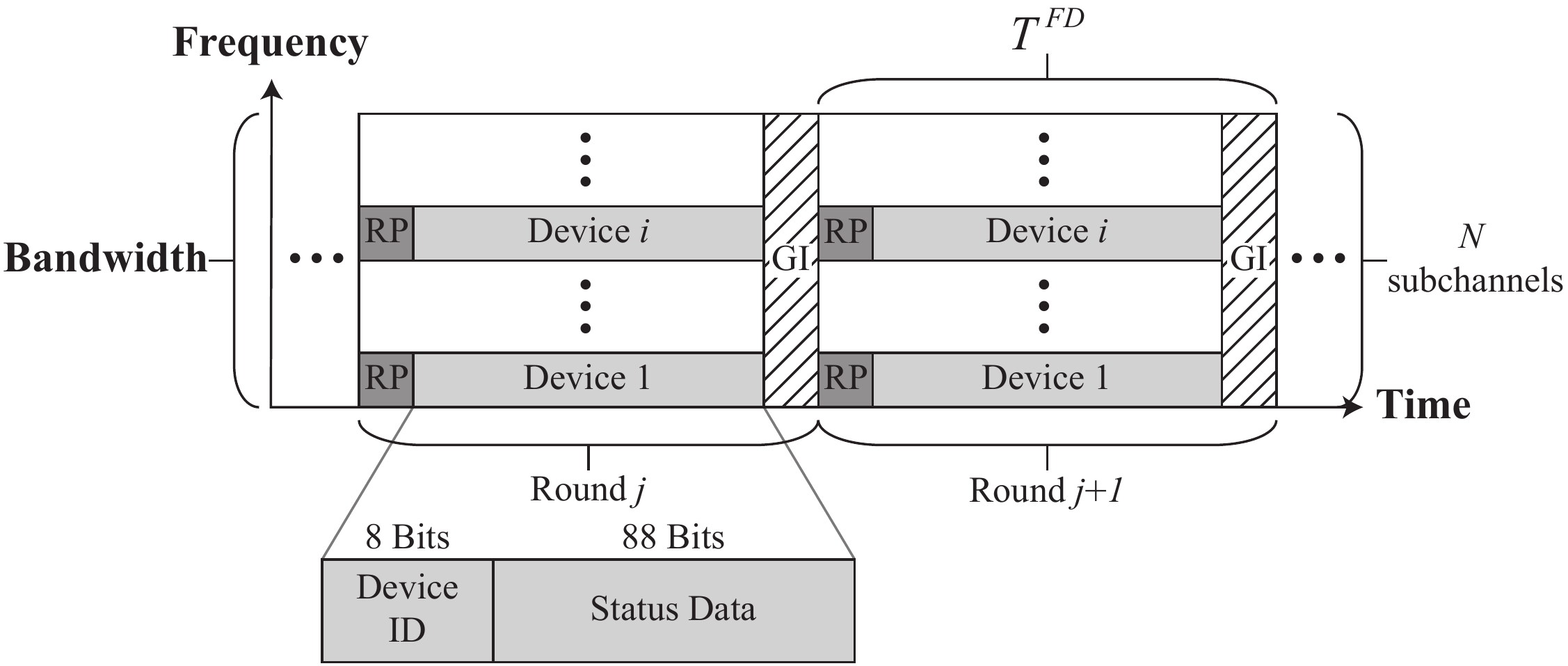}
    \caption{ The MAC protocol of the FDMA information update system.}
    \label{fig:fdma_frame}
\end{figure}

\begin{table}[t]
	\caption{\textcolor{black}{The Subcarrier Allocation in the FDMA System.}}\label{tab:fdma_subc_alloc}
	\centering
	\begin{tabular}{c|c}
		\hline
		\textbf{Users} & \textbf{Subcarriers} \\ \hline
		$1$  & $9-16$ \\ \hline
		$2$  & $17-24$ \\ \hline
		$3$  & $25-32$ \\ \hline
		$4$  & $34-41$ \\ \hline
		$5$  & $42-49$ \\ \hline
		$6$  & $50-57$ \\ \hline
	\end{tabular}
\end{table}

The MAC protocol of the FDMA information update system is shown in Fig. \ref{fig:fdma_frame}. As in the TDMA systems, each update packet is preceded by a reduced preamble before the payload. In the status update packet, $8$ bits are used for the device ID, and $88$ bits are used for the status data. These bits are then channel-encoded into $192$ coded bits. Note that since the devices in the FDMA system constantly generate and send new status after a transmission round, the AP does not need to send an ACK to notify the status of the previous transmission.\footnote{We also omit other control frames such as synchronization frames because they do not need to be transmitted frequently in our time-synchronized multiuser OFDM systems. The implementation details for synchronization can be found in \cite{liang_design_2021}.}

Note that the OFDMA implementation in \cite{liang_rofa_2021} eliminates the STS from the device transmit packets, leaving only the LTS for channel estimation. We left a GI after the status packet to avoid inter-packet interference due to propagation delays. In this setup, the duration of an FDMA time slot ${T^{FD}}$ can be calculated by
\begin{equation}
    {T^{FD}} = \left( {160 + \frac{{192}}{8} \times 80} \right) \times {10^{ - 4}} + 0.016 = 0.224\text{ ms}.
\end{equation}

\subsubsection{Experimental Results}
We conduct USRP SDR experiments in both power-balanced and power-imbalanced cases. We control the target received signal power at the AP by adjusting the transmit power on the transmitter side. During the experiment, we vary the SNR of the devices and run ${10^5}$ frames for each SNR value to examine the average AoC performance of the system. The experimental results for the power-balanced and power-imbalanced cases are shown as follows.

\begin{figure}[t]
    \centering
    \includegraphics[width=3.5in]{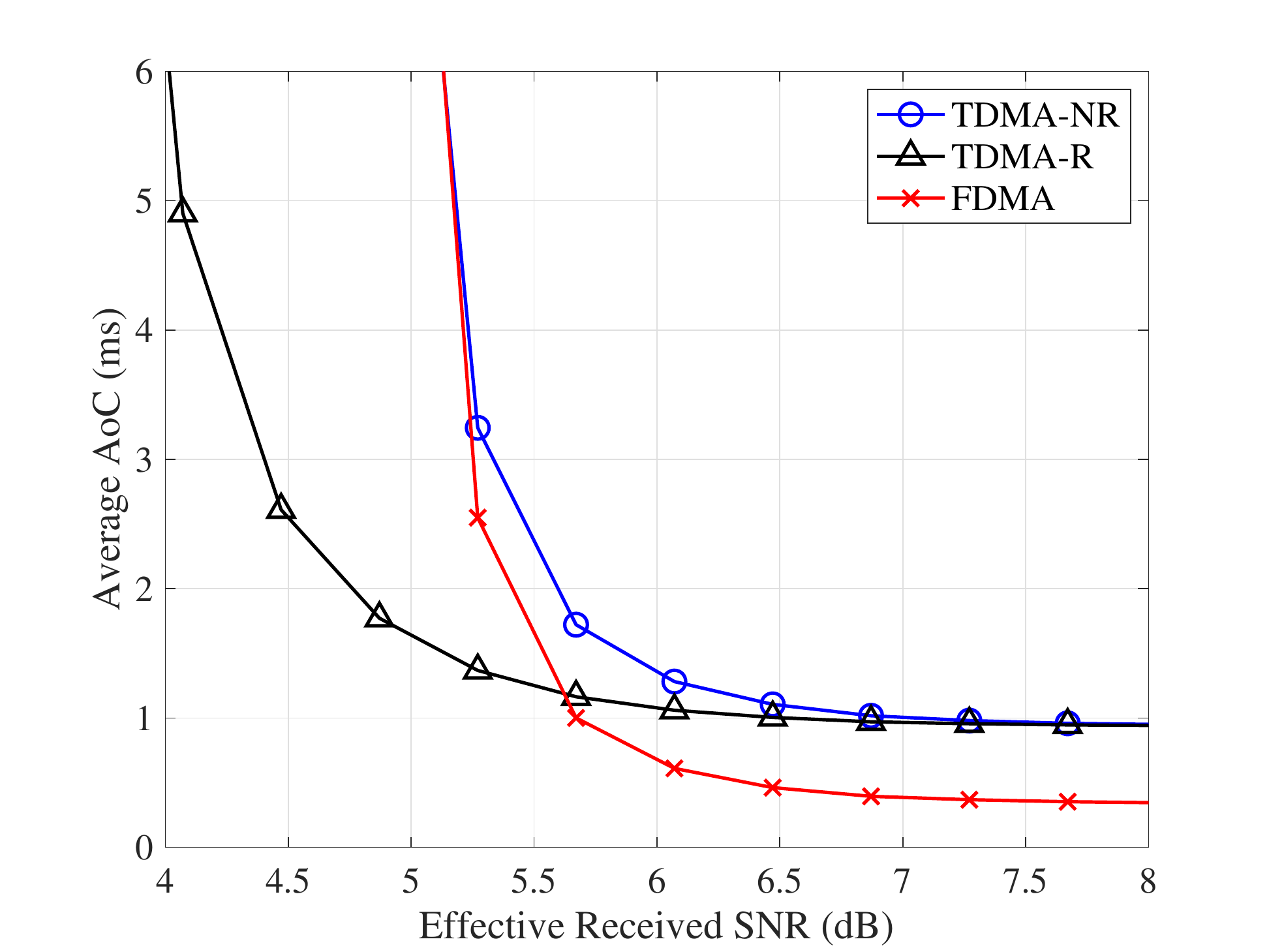}
    \caption{The average AoC of the $6$-user power-balanced information update system using the three multiple access schemes in the USRP SDR experiments.}
    \label{fig:exp_aoc_balanced}
\end{figure}

\paragraph{Power-balanced Case}
In the power-balanced case, the average effective received power of a device signal at the AP is approximately equal to that of the other devices. The experimental results of the AoC performance of the power-balanced systems are depicted in Fig. \ref{fig:exp_aoc_balanced}. We use the average AoC in milliseconds as the y-axis in the figure.

Fig. \ref{fig:exp_aoc_balanced} shows that TDMA-R outperforms both TDMA-NR and FDMA in the low SNR regime and also outperforms TDMA-NR in the middle SNR regime in terms of average AoC. Meanwhile, TDMA-NR and FDMA have close performance in the low SNR regime. Nevertheless, it is noteworthy that FDMA outperforms both TDMA schemes in the high SNR regime. Specifically, the minimum average AoC of FDMA is less than $50\%$ of the minimum average AoC of TDMA.

Recall that in the theoretical results, TDMA-R achieves overall better performance than the other two multiple access schemes. The conclusion of the experimental results differs from the theoretical results because the ACK transmission and GI overhead are considered in the experiments. When the SNR is low, the reduction in AoC from preventing unnecessary retransmissions of packets successfully delivered in TDMA-R is significant. However, when the SNR is high, the packet success rate is also high; thus, all schemes require fewer retransmissions. As a result, the reduction in AoC comes from reducing ACK transmission and GI overhead in FDMA becomes dominant. While FDMA performs better than TDMA-R in the high SNR regime, TDMA-R has the most stable average AoC performance among the three schemes, which is critical for IIoT applications. The dramatic increase in average AoC of FDMA in the low SNR regime is undesirable for IIoT applications that require high stability.

\paragraph{Power-imbalanced Case}
In time-varying wireless environments, devices are likely to have different channel conditions with respect to the AP. Therefore, we now further investigate the average AoC of the three multiple access schemes in the practical power-imbalanced case. In the USRP SDR experiments, we let device $1$ be the strongest user with the highest effective received SNR at the AP. Devices $2$ to $5$ have (almost) the same effective received SNR at the AP, which is about $2$ dB lower than device $1$. Device $6$ has the lowest effective received SNR at the AP, which is about $4$ dB lower than device $1$. In this subsection, we consider that the user with the lowest SNR (i.e., user $6$) transmits first in each transmission round of TDMA-R and TDMA-NR.

\begin{figure}[t]
    \centering
    \includegraphics[width=3.5in]{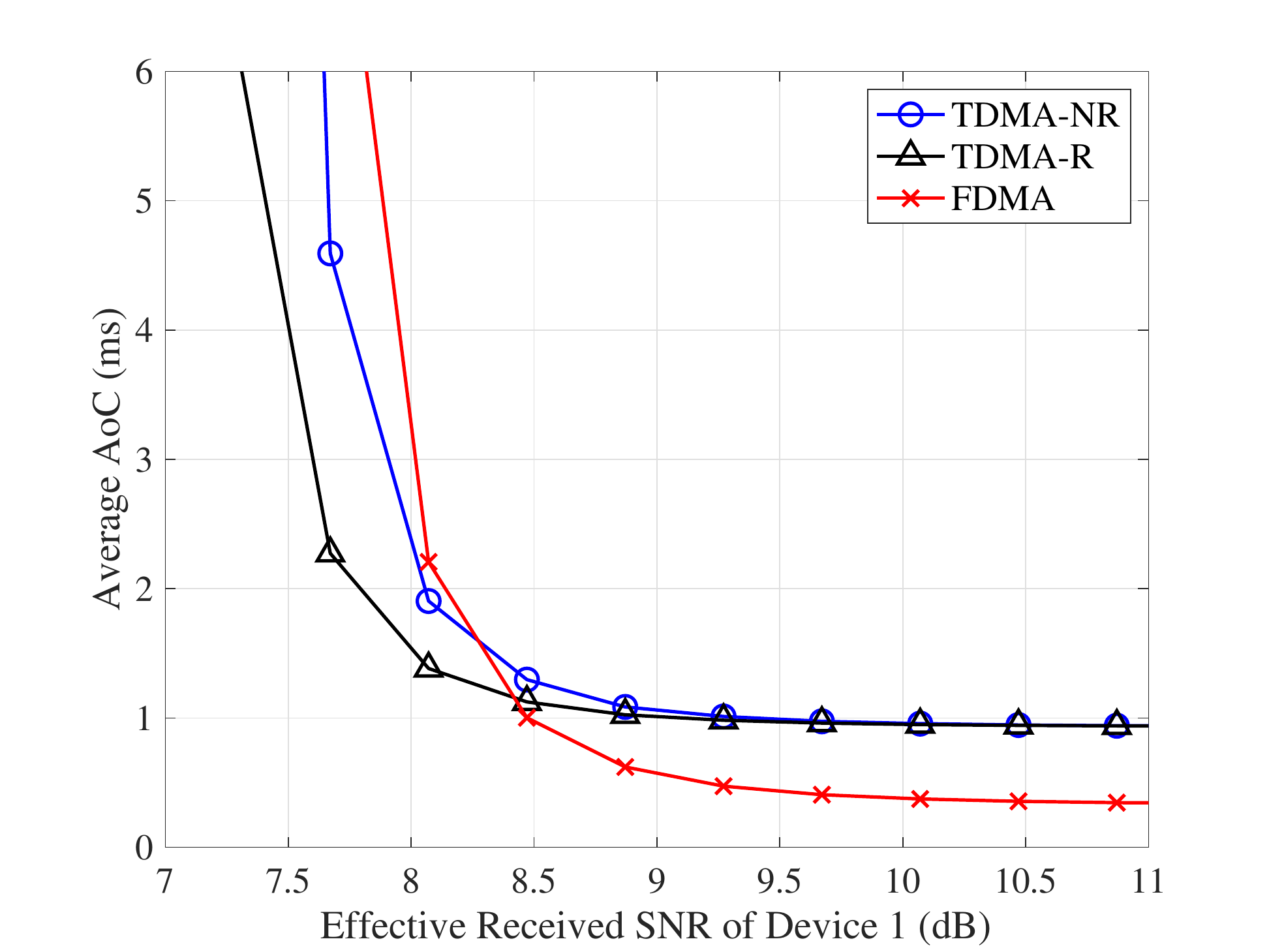}
    \caption{The average AoC of the $6$-user power-imbalanced system adopting the three schemes in the USRP SDR experiments.}
    \label{fig:exp_aoc_imbalanced}
\end{figure}

Fig. \ref{fig:exp_aoc_imbalanced} shows the average AoC of the three schemes with the same power-imbalanced setting. We use the effective received SNR of device $1$ (the strongest user) as the x-axis. The experimental results are generally consistent with those in the power-balanced case. Among the three schemes, TDMA-R achieves the best AoC performance in the low SNR regime, while FDMA performs the best in the high SNR regime. 

TDMA-R does not generate unnecessary retransmissions of successfully transmitted packets, resulting in a relatively low average AoC in all SNR regimes. When the SNR becomes high, the significant overhead caused by ACK transmissions and GIs in TDMA schemes makes FDMA a better choice, since all schemes require fewer retransmissions. However, TDMA-R exhibits the most stable AoC performance over a wide range of SNRs. Therefore, in practical wireless environments with varying SNRs, TDMA-R is a viable solution to provide both stable and low average AoC.

\paragraph{Effect of Transmission Order in TDMA}
In our experiments, we find that the transmission order of TDMA-NR and TDMA-R affects the AoC performance of the systems, as packets from IoT devices with lower SNRs at the AP are more likely to fail to be decoded and trigger a new transmission round or packet retransmission. Note that the devices in FDMA always transmit simultaneously; hence there is no transmission order in FDMA.

\begin{figure}[t]
    \centering
    \includegraphics[width=3.5in]{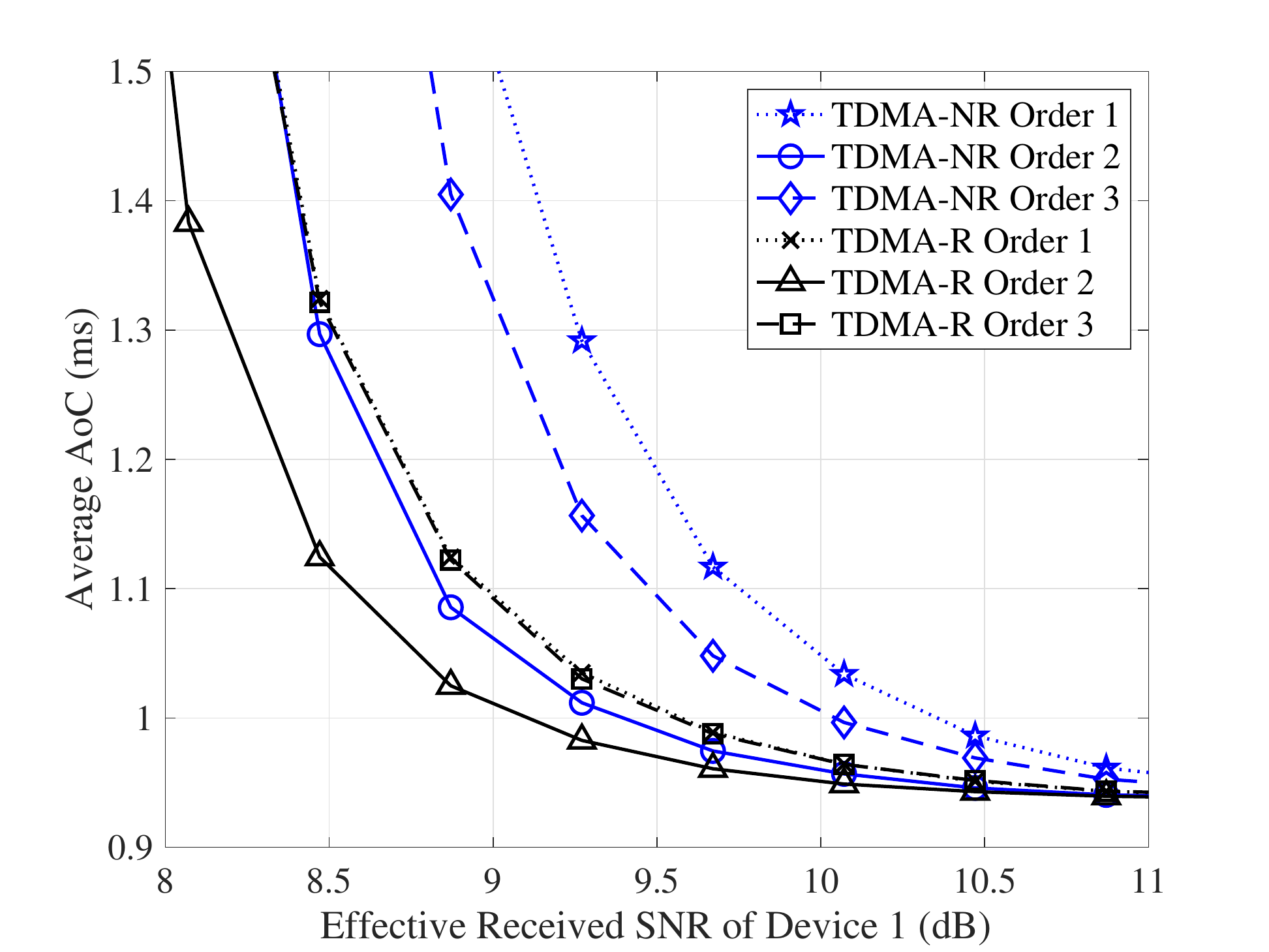}
    \caption{The average AoC of the two TDMA schemes using three different transmission orders.}
    \label{fig:exp_aoc_order}
\end{figure}

We conduct USRP SDR experiments to compare three transmission orders for TDMA-R and TDMA-NR. The transmission orders are given in Table \ref{tab:transmission_order}. These three transmission orders represent the following three cases: (1) the weakest device (device 6) transmits at the end, (2) the weakest device transmits at the beginning, and (3) the weakest device transmits in the middle. Fig. \ref{fig:exp_aoc_order} shows the average AoC of TDMA-NR and TDMA-R using the three transmission orders. From the results, we can conclude that the transmission order influences the average AoC of both TDMA-NR and TDMA-R systems.

Specifically, for the TDMA-NR system, the three transmission orders result in different AoC performances, and order $2$ provides the lowest average AoC among the three orders. This is because the earlier the weakest device transmits, the fewer status packets successfully transmitted by the other devices are wasted. In contrast, for the TDMA-R system, order $2$ exhibits the lowest average AoC, while orders $1$ and $3$ have close performance. Recall that if the transmission of the first device in TDMA-R fails, all the devices will sample the latest status in the subsequent time slot. Letting the weakest device transmit first prevents successfully transmitted status packets from becoming unnecessarily old due to retransmissions from the weakest device. With the first device in the transmission round remaining unchanged, the different transmission orders (i.e., orders $1$ and $3$) do not affect the average AoC, since TDMA-R allows immediate retransmissions. In other words, TDMA-R has a more stable AoC performance under different transmission orders compared to TDMA-NR. This facilitates the AP to pre-schedule the transmission of devices, which is advantageous in IIoT communications.

\begin{table}[t]
	\caption{\textcolor{black}{The Transmission Order of Devices in the TDMA Systems.}}\label{tab:transmission_order}
	\centering
	\begin{tabular}{c|c}
		\hline
		\textbf{Transmission Order} & \textbf{Devices} \\ \hline
		Order $1$  & $1,2,3,4,5,6$ \\ \hline
		Order $2$  & $6,1,2,3,4,5$ \\ \hline
		Order $3$  & $1,2,3,6,4,5$ \\ \hline
	\end{tabular}
\end{table}

\section{Conclusion}
We analyzed the AoC performance of different multiple access schemes: TDMA-NR, TDMA-R, and FDMA. Specifically, we focused on the situation where multiple IoT devices cooperated for a joint observation, with each device responsible for a part of the monitoring. Unlike the AoI metric, AoC is a new performance metric that measures the time elapsed since the last successful collection of all updates for the common observation. Due to the new rule for age updates, determining how to strategically divide the time-frequency resources for multiple access communication to achieve a low average AoC remains an open question. 

To address this issue, we derived the theoretical time-averaged AoC of the three multiple access schemes. Our theoretical analysis shows that TDMA-R has better average AoC performance than TDMA-NR and FDMA, especially when the SNR is low. This is because TDMA-R can immediately retransmit the packet without spending time waiting for the complete transmission of other cooperative packets. Moreover, we conducted experiments on the USRP SDR platform to compare the practical performance of these three schemes with the MAC protocols considered. Interestingly, our experimental results show that FDMA can perform better than TDMA-R in the high SNR regime due to the significant overhead caused by ACK transmission and GI in TDMA-R. Nevertheless, TDMA-R exhibits the most stable AoC performance over a wide range of SNRs, indicating that TDMA-R is a viable solution in practical wireless environments with varying SNRs.


%

\appendices




\ifCLASSOPTIONcaptionsoff
  \newpage
\fi

\end{document}